\documentclass[letter,11pt]{article}
\pdfoutput=1 
\usepackage{jcappub} 
\usepackage[normalem]{ulem}
\usepackage[utf8]{inputenc}
\usepackage[T1]{fontenc} 
\usepackage{dsfont}
\usepackage{amsmath,amssymb,calc}
\usepackage{color}
\usepackage{amssymb}
\usepackage{graphicx, epsfig, bm}
\usepackage{hyperref}
\usepackage{soul}
\usepackage{multicol}
\usepackage{changepage}
\usepackage{subfig}
\usepackage{array} 
\usepackage{mathtools}
\usepackage{comment}
\usepackage{color}
\usepackage{array}
\usepackage{ifthen}
\allowdisplaybreaks

\def\bdm{\begin{displaymath}}
	\def\edm{\end{displaymath}}

\def\barray{\begin{array}}
	\def\earray{\end{array}}
\def\be{\begin{equation}}
	\def\ee{\end{equation}}
\def\ben{\begin{equation} \nonumber}
	\def\een{\end{equation}}
\def\ban{\begin{eqnarray*}}
	\def\ean{\end{eqnarray*}}
\def\ba{\begin{eqnarray}}
	\def\ea{\end{eqnarray}}
\def\eal{\end{align}}
\def\bal{\begin{align}}

\def\({\left(}
\def\){\right)}
\def\[{\left[}
\def\]{\right]}

\def\by{{\bf{y}}}
\def\ba{{a\left(\by\right)}}

\def\bk{{\bf k}}
\def\bq{{\bf q}}
\def\br{{\bf r}}
\def\bs{{\bf s}}
\def\bx{{\bf x}}
\def\bp{{\bf p}}

\definecolor{gold}{rgb}{1.0, 0.84, 0.0}
\definecolor{maroon}{rgb}{.25,0,0}
\definecolor{darkorange}{rgb}{1.0, 0.55, 0.0}
\definecolor{corn}{rgb}{0.98, 0.93, 0.36}
\definecolor{bronze}{rgb}{0.8, 0.5, 0.2}
\definecolor{darkgreen}{cmyk}{0.85,0.2,1.00,0.2}

\begin{document}

	\title{Gravitational wave anisotropies from axion inflation}
	
	\author{Sofia P. Corb\`a}
	\affiliation{Amherst Center for Fundamental Interactions, Department of Physics, University of Massachusetts,
		1126 Lederle Graduate Research Tower, Amherst, MA 01003-9337 U.S.A.}
	
	
	
	\emailAdd{spcorba@umass.edu}
	
\abstract{An important prediction of inflation is the production of a primordial stochastic gravitational wave background. Observing this background is challenging due to the weakness of the signal and the simultaneous presence of an astrophysical background generated by many unresolved late-time sources. One possible way to distinguish between the two is to examine their anisotropies. In this paper we calculate the primordial correlation function of gravitational wave anisotropies in the cosmological background generated by axion inflation, where the inflaton is a pseudo-Nambu–Goldstone boson coupled to gauge fields. In this scenario, tensor modes arise not only from the standard amplification of vacuum fluctuations present in any inflationary model, but also from the inverse decay process of the produced gauge fields. The correlator of gravitational wave anisotropies consists therefore of two main components: the contribution from vacuum tensor modes and the contribution from tensor modes sourced by the gauge fields. Our analysis shows that, while the former, previously studied in the literature, is negligible, the one arising from the sourced tensor modes, normalized by the fractional energy density at interferometer frequencies, can reach values as large as $\mathcal{O}(10^{-1})$. This result shows that axion inflation can generate large anisotropies with the potential to be observed by gravitational wave detectors within a reasonable time frame.}
	
	\maketitle
	
	
	\section{Introduction}%
	
	Gravitational waves (GW) have recently received a lot of attention, especially after their first detection in September 2015 by the LIGO/Virgo collaboration~\cite{LIGOScientific:2016aoc} and the more recent evidence for a stochastic gravitational wave background (SGWB) reported by  pulsar timing array (PTA) measurements~\cite{NANOGrav:2023gor,EPTA:2023fyk,Reardon:2023gzh}. This background could be either astrophysical (AGWB), generated from unresolved astrophysical sources in later epochs, or cosmological (CGWB), which originates from phenomena in the early Universe such as inflation, reheating, phase transitions, primordial black holes, or topological defects~\cite{Allen:1996vm,Caprini:2018mtu,Regimbau:2011rp}. Investigating the CGWB provides information about the dynamics prevalent at the time of generation of the primordial GWs, offering a unique window into the physics of the early Universe.
	
	In this paper we focus on cosmological gravitational waves originated from a particular inflationary model known as axion inflation~\cite{Pajer:2013fsa}. In axion inflation, the inflaton is a pseudo-Nambu-Goldstone boson exhibiting a broken shift symmetry, i.e., a symmetry under the transformation $\phi\rightarrow \phi \,+\, \text{const.}$, which protects the flatness of the potential against large radiative corrections. In this model, proposed for the first time in 1990 as natural inflation~\cite{Freese:1990rb}, the inflaton interacts with gauge fields through the coupling $\phi F_{\mu\nu}\tilde{F}^{\mu\nu}/f$, where $f$ is the axion decay constant. As a consequence, the gauge field quanta get amplified and in turn produce scalar and tensor fluctuations through a process of inverse decay. Therefore, in axion inflation, both scalar fluctuations and  gravitational waves are generated through two distinct mechanisms: first, from the vacuum, via the standard amplification process, and second, as a consequence of the production of gauge fields, through an inverse decay process. Remarkably, because of the parity-violating nature of the system, only photons of a given helicity are produced~\cite{Sorbo:2011rz}, implying that the sourced gravitational waves of different helicities have different amplitudes. 
	
	The phenomenological predictions of axion inflation are multiple, including nongaussianities~\cite{Barnaby:2010vf}, deviations from scale invariance~\cite{Namba:2015gja}, formation of primordial black holes~\cite{Linde:2012bt}, baryogenesis~\cite{Anber:2015yca}, generation of cosmologically relevant magnetic fields~\cite{Garretson:1992vt, Anber:2006xt}, as well as generation of primordial chiral gravitational waves at CMB~\cite{Sorbo:2011rz} or interferometer~\cite{Cook:2011hg} frequencies. In particular, we expect these gravitational waves to generate an SGWB of cosmological origin, the characterization of which is essential for distinguishing it  from its astrophysical counterpart. 
	
	One method for characterizing the SGWB involves examining its anisotropies. In fact, the SGWB is expected to present small spatial fluctuations analogous to the temperature fluctuations of the CMB, the detection of which is a major challenge for the next generation of gravitational wave detectors ~\cite{KAGRA:2021mth,LISACosmologyWorkingGroup:2022kbp}. More importantly, these anisotropies may correlate with those of the CMB and the study of this cross-correlation provides a powerful way to distinguish between astrophysical and cosmological origins of the background~\cite{Geller:2018mwu,Malhotra:2020ket,Adshead:2020bji,Ricciardone:2021kel,Braglia:2021fxn,Schulze:2023ich}.
	
	In the specific context of axion inflation, reference~\cite{Corba_2024} analyzed the correlation between the curvature perturbation $\zeta(\bx)$ and the gravitational energy density $\Omega_{GW}(\bx)= \dot{h}_{ij}(\bx)\,\dot{h}_{ij}(\bx)/(12\,H_0^2)$. In axion inflation, both scalar fluctuations and gravitational waves have vacuum and sourced contributions.  At the same time, the expansion of the Universe induces vacuum fluctuations in the inflaton, leading to spatial variations in the gauge field population, which in turn generate spatial fluctuations in the sourced gravitational waves. As a result, the sourced gravitational waves consist of two components: one that we denote as the \textit{homogeneous} component, and the other as the component of \textit{fluctuations}. The homogeneous component arises from the gauge field and depends on the zero mode of the rolling inflaton. In contrast, the fluctuations originate from the gauge field's inhomogeneities, which are, in turn, imprinted by the inflaton's fluctuations.

	The correlator studied in~\cite{Corba_2024} receives two contributions: one from the correlation of the sourced gravitational waves with the vacuum scalar fluctuations, and the other from the correlation of the sourced gravitational waves with the sourced scalar fluctuations. The former effect is generally dominant and the correlator, normalized by the amplitude of $\zeta$ and by the fractional energy in sourced gravitational waves at interferometer frequencies, turned out to be of the order of $ 10^{-4}\div 10^{-2}$. The observability of this correlation, influenced by the intrinsic variance of the isotropic component and instrumental noise~\cite{Mentasti:2023icu,Cui:2023dlo}, depends not only on the overall gravitational wave energy density, but also on the amplitude of anisotropies in the gravitational wave spectra. Studies on preheating at the end of inflation and on baryogenesis suggest that these anisotropies may be large~\cite{Bethke:2013aba,Yu:2025jgx}. 
	
	In this work, we investigate the anisotropies in the gravitational wave spectra produced during axion inflation by computing the correlator $\langle\Omega_{\rm GW}(\bx)\Omega_{\rm GW}(\by)\rangle$ of the gravitational wave energy densities. This correlator consists of two main contributions: one arising from the correlation of gravitational wave energy densities generated by the vacuum tensor modes, which we refer to as the \textit{vacuum correlator}, and the other from the correlation of energy densities associated with the sourced tensor modes, called \textit{sourced correlator}. Since the vacuum correlator has already been studied in the literature~\cite{LISACosmologyWorkingGroup:2022kbp, Bartolo:2019oiq,Bartolo:2019yeu}, we will only present it briefly using an analytical approach, and instead focus primarily on the sourced correlator.

	The sourced correlator arises from three distinct contributions,  reflecting the fact that the sourced gravitational waves are composed of a homogeneous component and fluctuations. The first contribution comes from the correlation of the homogeneous components, resulting in the \textit{intrinsic correlator}. The second contribution comes from the correlation between the homogeneous components and the fluctuations, referred to as the  \textit{extrinsic correlator}. Finally, the third contribution arises from the correlation of the fluctuations and represents a higher-order contribution in the perturbations.
	
    Our analysis shows that the sourced extrinsic correlator, normalized by the square of the fractional energy in sourced gravitational waves at interferometer frequencies,  lies in the range $\mathcal{O}(10^{-5} - 10^{-1})$. In contrast, the sourced intrinsic correlator is significantly smaller, while the correlator of the sourced fluctuations is negligible. The vacuum correlator is also found to be small and unobservable. The relatively large value of the sourced extrinsic correlator,  which is the main result of this paper, is particularly significant, as it implies that the resulting anisotropies lie within the observational reach of GW detectors.
	
	The sourced gravitational waves studied in the sourced correlator are produced towards the end of axion inflation, when the amplitude of the gauge fields becomes large and they significantly backreact on the background inflationary evolution. Although the inflaton-gauge field dynamics is nonperturbative~\cite{Cheng:2015oqa, Notari:2016npn,Sobol:2019xls,DallAgata:2019yrr,Domcke:2020zez,Caravano:2022epk,Peloso:2022ovc,Figueroa:2023oxc,Garcia-Bellido:2023ser,vonEckardstein:2023gwk,Caravano:2024xsb}, the production of gravitational waves can be considered at the perturbative level. In~\cite{Garcia-Bellido:2023ser}, the authors showed that although backreaction can modify the dynamics of the system, the behavior of the sourced gravitational waves depends only on the velocity of the inflaton field, assuming inflaton inhomogeneities are neglected. Since our results are expressed entirely in terms of $\dot\phi(t)$, we assume them to remain valid even in the strong backreaction regime.
	
	This paper is organized as follows. In Section~\ref{sec:background}, we review the model of axion inflation, explaining the mechanism of gauge field amplification and the resulting production of scalar fluctuations (Subsection~\ref{Scalar fluctuations}) and gravitational waves (Subsection~\ref{Tensor fluctuations}). 
	In Section~\ref{Sec:correlator}, we define the correlator of the gravitational wave energy densities. In Section~\ref{sec: Sourced correlator}, we present the sourced correlator,  while Section~\ref{sec: Vacuum correlator} provides a brief overview of the vacuum correlator. Finally, in Section~\ref{sec:conclusion} we discuss our results and we conclude.

	\section{Overview of the axion inflation model}
	\label{sec:background}

	The action which describes our model of axion inflation is that of a pseudoscalar inflaton field $\phi$  minimally coupled to gravity and to a $U(1)$ gauge field $A_\mu$
	\begin{align}
		{\cal S}=\int d^4x\sqrt{-g}\left[\frac{M_P^2}{2}R-\frac12\partial_\mu\phi\,\partial^\mu\phi-V(\phi)-\frac14F_{\mu\nu}\,F^{\mu\nu}-\frac{\phi}{8\,f}\frac{\epsilon^{\mu\nu\rho\lambda}}{\sqrt{-g}}F_{\mu\nu}\,F_{\rho\lambda}\right]\,, 
	\end{align}
	where $g={\rm{det}}(g_{\mu\nu})$, $R$ is the Ricci scalar, $F_{\mu\nu}=\partial_\mu A_\nu-\partial_\nu A_\mu$ is the gauge field strength, $\epsilon^{\mu\nu\rho\lambda}$ is the totally antisymmetric tensor defined by $\epsilon^{0123}=+1$, $f$ is the axion decay constant and $V(\phi)$ is a generic inflationary potential.
	
	The quantum scalar and tensor fluctuations produced during inflation are obtained by adding spatially varying perturbations to the inflaton and the metric, respectively. In particular, the curvature perturbation $\zeta\equiv - \frac{H}{\dot{\phi}_0}\delta\phi$, where the overdot denotes the derivative with respect to cosmic time 
	$t$ (in contrast to the prime, which denotes the derivative with respect to the conformal time $\tau$), is related to the inflaton perturbations arising from
	\begin{align}
		\phi(\bx,\,\tau)=\phi_0(\tau)+\delta\phi(\bx,\,\tau)\,.     	\end{align}
	Gravitational waves, on the other hand, are obtained by introducing spatially varying perturbations in the form of transverse traceless tensor modes, i.e. $h_{ij}(\bx,t)$ with $h_{ii}=\partial_ih_{ij}=0$, to the de Sitter metric
	\begin{align}\label{eq:metric}
		ds^2&=a^2(\tau)\left[-d\tau^2+\left(\delta_{ij}+h_{ij}(\bx,\,\tau)\right)\,dx^i\,dx^j\right]\,,
	\end{align}
	where for repeated latin indices Einstein notation is used. The scale factor is $a(\tau)=-1/(H\,\tau)$, and it is set to be equal to unity at the end of inflation, i.e., at $\tau_{e}=-1/H$.
	
	To proceed, we expand the Lagrangian density around the background solution, identified by  $\phi_0(\tau)$ and $a(\tau)$, and then discard the terms of zeroth and first order in the perturbations.  By choosing the Coulomb gauge, i.e. $A_0(\bx,\,\tau)=0$ and $\partial_i A_i(\bx,\,\tau)=0$, the perturbed Lagrangian takes the form
	\begin{align}\label{eq:quad_lag}
		{\cal L}&=\frac{1}{2}\Phi'{}^2-\frac{1}{2}\partial_k\Phi\,\partial_k\Phi+\frac{a''}{2\,a}\Phi^2+\frac12{H}'_{ij}\,{H}'_{ij}-\frac12\partial_k{H}_{ij}\,\partial_k{H}_{ij}+\frac{a''}{2\,a}{H}_{ij}\,{H}_{ij}+\frac12A_i'\,A_i'\nonumber\\
		&-\frac12\partial_kA_i\,\partial_kA_i-\frac{\phi_0}{f}\epsilon^{ijk}\,A_i'\,\partial_jA_k-\frac{H_{ij}}{a\,M_P}\,\left[A_i'\,A_j'-\left(\partial_iA_k-\partial_kA_i\right)\left(\partial_jA_k-\partial_kA_j\right)\right]\nonumber\\
		&-\frac{\Phi}{f\,a}\epsilon^{ijk}\,A_i'\,\partial_jA_k\,,
	\end{align}
	where we have expressed the scalar and tensor perturbations  $\delta\phi$ and $h_{ij}$ in terms of their canonically normalized versions
	\begin{align}\label{eq:can_norm_quantities}
		&\Phi(\bx,\,\tau)= a(\tau)\,\delta\phi(\bx,\,\tau)\,,\nonumber\\
		&H_{ij}(\bx,\,\tau)=\frac{M_P}{2}\,a(\tau)\,h_{ij}(\bx,\,\tau)\,.
	\end{align}

By varying the Lagrangian \eqref{eq:quad_lag} with respect to  $A_i$, $\Phi$ and $H_{ij}$, we obtain the equations of motion that govern the dynamics of the system
	\begin{align}	\label{eq:eom_A}
		&A_i'' -\nabla^2 A_i -\frac{\phi'_0}{f}\epsilon^{ijk}\,\partial_jA_k=0\,,\\ \label{eq:eom_Phi}
		&\Phi'' - \frac{a''}{a}\Phi- \nabla^2\Phi+\frac{1}{f\,a}  \epsilon^{ijk}\,A_i'\,\partial_jA_k=0\,,\\
		\label{eq:eom_H}
		&H''_{ij}- \frac{a''}{a} H_{ij} -\nabla^2 H_{ij} +\frac{1}{a\,M_P}\,\left[A_i'\,A_j'-\left(\partial_iA_k-\partial_kA_i\right)\left(\partial_jA_k-\partial_kA_j\right)\right]\,=0\,.
	\end{align}

	Equation~\eqref{eq:eom_A} describes the evolution of the gauge fields. To study the amplification of gauge modes due to the rolling inflaton, we promote the classical field to an operator $\hat{A}_i(\bx,\,\tau)$, which we decompose into creation/annihilation operators $\hat{a}^\dagger_\lambda(\bk)/\hat{a}_{\lambda}(\bk)$, satisfying  the usual commutation relations $[\hat{a}_{\lambda}(\bk),\,\hat{a}_{\lambda'}^\dagger(\bq)]=\delta(\bk-\bq)\,\delta_{\lambda,\lambda'}\,$, $[\hat{a}_{\lambda}(\bk),\,\hat{a}_{\lambda'}(\bq)]=[\hat{a}_{\lambda}^\dagger(\bk),\,\hat{a}_{\lambda'}^\dagger(\bq)]=0$\,,
	\begin{align}\label{eq:quantA}
		\hat{A}_{i}(\bx,\,\tau)&=\int\frac{d\bk}{(2\pi)^{3/2}}\sum_{\lambda=\pm} e^{\,\lambda}_{i}(\widehat\bk)\,e^{i\bk\bx}\left[A_\lambda(k,\,\tau)\,\hat{a}_\lambda(\bk)+A_\lambda^*(k,\,\tau)\,\hat{a}^\dagger_\lambda(-\bk)\right]\,,
	\end{align}
	with the helicity projectors $e^{\lambda}_{i}(\widehat\bk)$ following the relations
	\begin{align}
		\begin{array}{ll}
			k_i\,e_i^{\,\lambda}(\widehat{\bk})=0\,,&e_i^{\,\lambda}(\widehat\bk)^*=e_i^{-\lambda}(\widehat\bk)=e_i^{\,\lambda}(-\widehat\bk)\,,\\
			i\epsilon_{ijk}k_je_k^{\,\lambda}(\widehat\bk)=\lambda\, k\, e_i^{\,\lambda}(\widehat\bk)\,,\qquad\qquad & e^{\,\lambda}_i(\widehat\bk)e_i^{\,\lambda'}(\widehat\bk)=\delta_{\lambda,\,-\lambda'}\,.
		\end{array}
	\end{align}
	Inserting eq.~\eqref{eq:quantA} into eq.~\eqref{eq:eom_A} and defining
	\begin{align}\label{xi}
		\xi\equiv \frac{\dot{\phi}_0}{2\,f\,H}\,,
	\end{align}
	we obtain the equation that governs the evolution of the mode functions $A_\lambda(k,\,\tau)$
	\begin{align}\label{eq:eomAlambda}
		\frac{d^2 A_\lambda(k,\,\tau)}{d\tau^2}+\left(k^2+\lambda\frac{2\,\xi\,k}{\tau}\right)\,A_\lambda(k,\,\tau)=0\,. 
	\end{align}
	Depending on the sign of $\xi$,  one of the two helicities, $\lambda=1$ or $\lambda=-1$, develops tachyonic instability. Assuming, without loss of generality, $\xi>0$ and keeping in mind that $\tau<0$ throughout the entire inflationary phase, the negative helicity mode $A_-$ has real frequencies that evolve adiabatically for all parameter values. As a result, $A_-$ remains in its vacuum state and can therefore be neglected. On the other hand, the positive helicity mode $A_+$ can acquire imaginary frequencies, leading to exponential amplification.
	
	More precisely, the solution of eq.~\eqref{eq:eomAlambda} that reduces to positive frequency as $k\tau \rightarrow -\infty$ can be explicitly expressed in terms of the regular and irregular Coulomb wave functions, $F_0$ and $G_0$, as $A_{\pm}= \frac{1}{\sqrt{2\,k}}(i\,F_0(\pm\xi, -k\tau)+G_0(\pm\xi, -k\tau))$. Under the WKB approximation, the leading term in the solution for the tachyonic modes of $A_+$, in the range $\frac{1}{8\,\xi}\lesssim |k\,\tau|\lesssim  2\,\xi$ ~\cite{Anber:2006xt, Barnaby:2010vf},  assuming~$\xi\gtrsim O(1)$ throughout, takes the form
	\begin{align}\label{eq:a+appr}
		A_+(k,\,\tau)\simeq \frac{1}{\sqrt{2\,k}}\,\left(-\frac{k\,\tau}{2\,\xi}\right)^{1/4}\,e^{-2\,\sqrt{-2\xi k\tau}+\pi\,\xi}\,,
	\end{align}
	which can be generalized to the entire range $0<|k\,\tau|<\infty$, as the observables of interest depend only on the range where the approximation is valid.	The positive helicity mode of the gauge field is therefore amplified by a factor of $e^{\pi\xi}$ and can become very large even for moderate values of $\xi$.
	
    The accelerated expansion of the background during axion inflation gives rise to the vacuum components of the scalar and tensor fluctuations, denoted respectively as $\delta\phi_{\rm V}$ and $h_{ij,\rm V}$, generated via the standard amplification process present in any inflationary model. The gauge fields amplified by the rolling zero mode of the inflaton are, on the other hand, responsible for the production of sourced scalar and tensor fluctuations through an inverse decay process, schematically represented as $\phi_0 \rightarrow A \rightarrow \{\delta\phi_{\rm S}, \,h^0_{ij,\rm S}\}$. Additionally, vacuum scalar fluctuations of the background inflaton induce fluctuations in the population of the produced gauge fields, resulting in fluctuations in the sourced tensor modes, schematically
	$\delta\phi_{\rm V} \rightarrow \delta A \rightarrow \delta h_{ij,\rm S}$. As a result of these mechanisms, analyzed in the next two Subsections, we obtain the fluctuations 
	\begin{align}
		\delta \phi = \delta \phi_{\rm V}+ \delta \phi_{\rm S} \hspace{0.5cm} \text{and} \hspace{0.5cm} 	h_{ij} = h_{ij, \rm V}+ h_{ij,\rm S}\,,
	\end{align}
	with 
	\begin{align}\label{dec_of_sourced_GW}
		h_{ij,\rm S} = h_{ij,\rm S}^0+ \delta h_{ij,\rm S}\,.
	\end{align}
The same decomposition holds also for the normalized versions of the fluctuations given in~\eqref{eq:can_norm_quantities}.
	\subsection{Scalar fluctuations}
	\label{Scalar fluctuations}
	The production of scalar fluctuations in the axion inflation model is described by eq.~\eqref{eq:eom_Phi}. The solution of the homogeneous part of the equation corresponds to the vacuum fluctuations $\Phi_{\rm V}$, generated as a result of the accelerated expansion of the background. This vacuum component can be quantized through the standard quantization of the free Lagrangian as
	\begin{align}\label{eq:quant_Phi}
		\hat{\Phi}_{\rm V}(\bx,\,\tau)&=\int\frac{d\bk}{(2\pi)^{3/2}}e^{i\bk\bx}\,\left[\Phi_{\rm V}(k,\,\tau)\,\hat{a}(\bk)+\Phi_{\rm V}^*(k,\,\tau)\,\hat{a}^\dagger(-\bk)\right]\,,
	\end{align}
	where 
	\begin{align}\label{eq:quant_Phi2}
		\Phi_{\rm V}(k,\,\tau)&\equiv\frac{1}{\sqrt{2k}}\left(1-\frac{i}{k\tau}\right)\,e^{-ik\tau}\,.
	\end{align}
	On the other hand, the particular solution of eq.~\eqref{eq:eom_Phi} corresponds to the sourced fluctuations, $ \Phi_{\rm S}$, produced by the amplified gauge fields. This solution, determined using the retarded propagator
	\begin{align}\label{eq:prop}
		G_k(\tau,\tau')= \frac{(1+k^2\,\tau\,\tau')\,\sin(k\,(\tau-\tau'))+k\,(\tau'-\tau)\,\cos(k\,(\tau-\tau'))}{k^3\,\tau\,\tau'}\,\Theta(\tau-\tau')\,,
	\end{align}
	where $\Theta$ denotes the Heaviside step function, is found to be
	\begin{align}\label{eq:phi_sourced}
		&\Phi_{\mathrm S}(\bq,\,\tau)\equiv i\,\int d\tau'\,G_q(\tau,\,\tau')\frac{H\tau'}{f}\,\epsilon^{ijk}\int\frac{d\bp}{(2\pi)^{3/2}} \,A_i'(\bp,\,\tau')\,(\bq-\bp)_jA_k(\bq-\bp,\,\tau')\,.
	\end{align}

	The complete solution is the sum of the two components, i.e.	$\Phi= \Phi_{\mathrm V}\,+\,\Phi_{\mathrm S}\,,$
	which gives rise to the curvature perturbation $	\zeta= \zeta_{\mathrm V}\,+\,\zeta_{\mathrm S} \,.$
The power spectrum of the curvature perturbation, ${\cal P}_\zeta$, defined through the two point function
	\begin{align}\label{two_point_function_curvature_perturbation}
		\langle \zeta(\bk)\,\zeta(\bq)\rangle= \frac{2\pi^2}{k^3}\,{\cal P}_\zeta(\bk)\,\delta(\bk+\bq)\,,
	\end{align}
	is the sum of the power spectra corresponding to the vacuum and sourced components, denoted as ${\cal P}_{\zeta,{\rm V}}$ and ${\cal P}_{\zeta,{\rm S}}$, respectively. 	The vacuum power spectrum associated with the mode functions in~(\ref{eq:quant_Phi2}), evaluated at the end of inflation and in the large scale limit, is given by
	\begin{align}
		{\cal P}_{\zeta,{\rm V}}=\frac{k^3}{2\pi^2}\frac{H^2}{\dot{\phi}_0^2}\left|\Phi_{\rm V}(k,\,\tau_{e})\right|^2\xrightarrow[k\ll H]{} \frac{H^4}{4\pi^2\,\dot{\phi}_0^2}\,,
	\end{align}
while, the sourced power spectrum corresponding to~(\ref{eq:phi_sourced}), for $\xi\gtrsim 3$, is found to be~\cite{Barnaby:2010vf}
	\begin{align}\label{sourced_power_spectrum}
		{\cal P}_{\zeta,{\rm S}}=\frac{k^3}{2\pi^2}\frac{H^2}{\dot{{\phi}}_0^2}\,\left|\Phi_{\rm S}(k,\,\tau_{e})\right|^2\xrightarrow[k\ll H]{} 4.8\times 10^{-8}\,\frac{H^8}{\dot\phi_0^4}\,\frac{e^{4\pi\xi}}{\xi^6}\,.
	\end{align}

		The CMB observations, particularly from the Planck satellite, have placed important constraints on nongaussianities at large scales, which are consistent with the predictions of single-field inflationary models. Specifically, the parameter $f_{\rm NL}$  used to quantify nongaussianity and defined through the three-point correlation function of the curvature perturbation
	\begin{align}
		\langle \zeta(\bk_1)\,\zeta(\bk_2)\,\zeta(\bk_3)\rangle=\frac{3}{10}\left(2\pi\right)^{5/2}\,f_{\rm NL}(k_1,\,k_2,\,k_3)\,{\cal P}_\zeta^2\,\delta(\bk_1+\bk_2+\bk_3)\,\frac{k_1^3+k_2^3+k_3^3}{k_1^3\,k_2^3\,k_3^3}\,,
	\end{align}
	in the context of single-field, slow-roll inflation, is predicted to be of the order of the slow-roll parameters~\cite{Maldacena:2002vr}.	In the model of axion inflation the sourced curvature perturbations can lead to large nongaussianities.  These perturbations arise from gauge fields through an inverse decay process, which maximizes the nongaussian effects in the equilateral configuration, i.e., when $k_1=k_2=k_3$. In this configuration, the nongaussianity parameter $f_{\rm NL}^{\rm equil}$ is given by~\cite{Barnaby:2010vf}
	\begin{align}\label{eq:fNL}
		f_{\rm NL}^{\rm equil}\simeq 7.1 \times 10^{5}\,\frac{H^{12}}{\dot\phi^6}\,\frac{e^{6\pi\xi}}{\xi^9}\,.
	\end{align}
	For large values of $\xi_{\rm CMB}$, i.e., the value of $\xi$ when CMB scales exit the horizon, the parameter $f_{\rm NL}^{\rm equil}$ exceeds the observational bounds on nongaussianity. In order to reproduce the observations it is  necessary that $\xi_{\rm CMB}\lesssim 2.5$~\cite{Barnaby:2011vw}. As a result, the sourced power spectrum~\eqref{sourced_power_spectrum} at this time is significantly suppressed and becomes subdominant compared to the vacuum contribution, i.e.  ${\cal P}_{\zeta,{\rm S}}\ll {\cal P}_{\zeta,{\rm V}}$. 
	The constraint on $ \xi $ imposes a restriction on the amplitude of the produced gauge field, which in turn must remain relatively small. Since the gauge fields are responsible for generating the sourced tensor modes, as described in the next Subsection, these modes will also be small at this stage.
	
	\subsection{Tensor fluctuations}
	\label{Tensor fluctuations}
	A similar analysis holds also for tensor fluctuations, which correspond to gravitational waves.  In this case, the homogeneous and particular solutions of eq.~\eqref{eq:eom_H} correspond, respectively, to the vacuum fluctuations $H_{ij,\rm V}$, generated by the expanding inflationary background, and the sourced fluctuations  $H_{ij,\rm S}$, produced through the inverse decay of the gauge fields.
	The vacuum component can be quantized once again through the standard quantization process, starting from the free Lagrangian as
	\begin{align}
		\hat{H}_{ij,\rm V} (\bx,\,\tau)&=\int\frac{d\bk}{(2\pi)^{3/2}}\sum_{\lambda=\pm} e^{\,\lambda}_{ij}(\widehat\bk)\,e^{i\bk\bx}\left[H_{\rm V}^{\lambda}(k,\,\tau)\,\hat{a}_\lambda(\bk)+{H_{\rm V}^{\lambda}}^*(k,\,\tau)\,\hat{a}^\dagger_\lambda(-\bk)\right]\,,
	\end{align}
	where 
	\begin{align}
		H_{\rm V}^{\pm}(k,\,\tau)&\equiv\frac{1}{\sqrt{2k}}\left(1-\frac{i}{k\tau}\right)\,e^{-ik\tau}\,,
	\end{align}
	and 
	\begin{align}e^{\,\lambda}_{ij}(\widehat\bk)= e^{\,\lambda}_{i}(\widehat\bk)\,e^{\,\lambda}_{j}(\widehat\bk).\end{align}
Using again the retarded propagator~\eqref{eq:prop} we find
	the particular solution as
	\begin{align}\label{eq:h_sourced}
		&H_{ij,\mathrm S}(\bq,\,\tau)\equiv \int d\tau' \,G_q(\tau,\tau')\, \frac{H \, \tau'}{M_P} \int \frac{d\bp}{(2\pi)^{3/2}} \,A_i'(\bp,\,\tau')\,A_j'(\bq-\bp,\,\tau')\,,
	\end{align}
which has been simplified by the fact that the electric field dominates over the magnetic field in strength. The complete solution is again the sum of the two components, i.e. $H_{ij}= H_{ij,\rm V}+H_{ij,\rm S}$, which give rise, respectively, to the vacuum and sourced power spectra~\cite{Sorbo:2011rz}
	\begin{align}\label{eq:Ph}
		{\cal P}_h={\cal P}_{h,{\rm V}}+{\cal P}_{h,{\rm S}}\simeq \frac{2\,H^2}{\pi^2\,M_P^2}+8.7\times 10^{-8}\,\frac{H^4}{M_P^4}\,\frac{e^{4\pi\xi}}{\xi^6}\,.
	\end{align}

		The scales relevant to the observation of gravitational waves, i.e., those measured by gravitational wave detectors, exit the horizon much later than the CMB scales, closer to the end of inflation. At these later times, the parameter $ \xi $, which typically increases (slowly) during inflation, reaches values larger than $ \xi_{\rm CMB} $. 
	Since the power spectrum of the sourced gravitational waves depends exponentially on $ \xi$, at these later times we can have 	$
	{\cal P}_{h,{\rm S}} > {\cal P}_{h,{\rm V}}
	$, in contrast to what occurs when the CMB modes leave the horizon. The exponential amplification of the sourced power spectrum makes it possible for the gravitational waves to be directly detectable by current or future GW detectors~\cite{Cook:2011hg}.
	
	The sourced gravitational waves produced through the mechanism described above are decomposed into a homogeneous part, which depends entirely on the background rolling inflaton 
	$\phi_0$, and fluctuations, which are caused by the vacuum fluctuations $\delta \phi_{\rm V}$, as described in the paragraph above eq.~\eqref{dec_of_sourced_GW}. To first order in the vacuum-amplified inflaton fluctuations $\delta\phi_{\rm V}$ and in the limit in which the wavelength of $\delta\phi_{\rm V}$ is much larger than that  of $h_{ij,{\rm S}}$, the sourced gravitational field decomposition in homogeneous part and fluctuations~\eqref{dec_of_sourced_GW} takes the form
	\begin{align}\label{hequalhzeroanddeltah}
		h_{ij, \rm S}(\bx,\,\phi(\bx))=h^0_{ij,\rm S}(\bx)+ \delta h_{ij,\rm S}(\bx)\,\equiv h_{ij,  \rm S}^0(\bx,\,\phi_0)+\frac{\partial h^0_{ij,\rm S}(\bx,\,\phi_0)}{\partial\phi_0}\,\delta\phi_{\rm V}(\bx) \,.
	\end{align}
	Since $h^0_{ij,{\rm S}}(\bx,\,\phi_0)\propto e^{2\pi\xi}$, $\delta h_{ij, \rm S}$ becomes
	\begin{align}\label{deltah}
		\delta h_{ij, \rm S} (\bx) = -2\pi\frac{d\,\xi}{d\phi_0}\,\frac{\dot\phi_0}{H}\,\zeta_{\rm V}(\bx)\,h^0_{ij, \rm S}(\bx)\,.
	\end{align}
	Assuming $\dot\phi_0>0$, $V'<0$, we have
	\begin{align}
		\xi\equiv\frac{\dot\phi_0}{2\,fH}\simeq -\frac{V'}{6\,fH^2}=-\frac{M_P^2}{2\,f}\frac{V'}{V}\,,
	\end{align}
	and using the usual slow roll parameters $	\epsilon=\frac{M_P^2}{2}\,\frac{V'{}^2}{V^2}$ and $\eta=M_P^2\frac{V''}{V}$, the derivative of $\xi$ with respect to the background inflaton becomes
	\begin{align}
		\frac{d\xi}{d\phi_0}=-\frac{M_P^2}{2\,f}\left(\frac{V''}{V}-\frac{V'{}^2}{V^2}\right)=\left(\epsilon-\frac{\eta}{2}\right)\frac{1}{f}\,.
	\end{align}
	Using again eq.~\eqref{xi} and considering that the parameter $\xi$ typically takes the values $2 \div 5$, and that the quantity $(2\epsilon-\eta)$ is of the order of $10^{-2}\div 10^{-1}$, we have
	\begin{align}\label{factor}
		2\pi\frac{d\,\xi}{d\phi_0}\,\frac{\dot\phi_0}{H} = 	2\pi \(2\, \epsilon-\eta\) \xi \,\sim \mathcal{O}\(.1\div 3\)\,.
	\end{align}

We can now study the anisotropies in the gravitational wave background produced in axion inflation by estimating the correlator of the gravitational wave energy densities. This correlator receives two contributions: the vacuum correlator, which corresponds to the gravitational wave energy densities of the vacuum tensor modes, and the sourced correlator, which corresponds to the energy densities of the sourced tensor modes. The sourced correlator can further be decomposed into three components: the intrinsic correlator, the extrinsic correlator and the correlator of the fluctuations, which however will be neglected being very small. In Section~\ref{Sec:correlator}, we present the general form of the correlator, while in Sections~\ref{sec: Sourced correlator} and~\ref{sec: Vacuum correlator}, we analyze the sourced and vacuum correlators, respectively.

	\section{The correlator of the gravitational wave energy densities}
		\label{Sec:correlator}
	The normalized correlator of the gravitational wave energy densities is defined as
\begin{align}\label{eq:CorrelatorgenericDefinition}
	{\cal C}_{\Omega \Omega}(\bk)&=\frac{1}{\Omega_{GW}^{\,2}}\frac{k^3}{2\pi^2}\int\,d\by\, e^{-i\bk\by}\,\langle \Omega_{GW}(\bx+\by,\,t_0)\,\Omega_{GW}(\bx,\,t_0)\rangle\,\nonumber\\ &=\frac{1}{\Omega_{GW}^{\,2}}\frac{k^3}{2\pi^2}\,\langle \Omega_{GW}(\bk,\,t_0)\,\Omega_{GW}(-\bk,\,t_0)\rangle'\,,
\end{align}
where $t_0$ is the present value of the cosmic time, 	$
	\Omega_{GW}\simeq\frac{\Omega^0_{\rm rad}}{24}\,{\cal P}_{h}(k_{\rm INT})$ with  $\Omega^0_{\rm rad} \simeq 8.2\times 10^{-5}$ is the fractional energy in gravitational waves at interferometer frequencies~\cite{Caprini:2018mtu} and $\langle \dots\rangle'$ represents the correlator stripped of the Dirac delta. Considering the explicit expression $\Omega_{GW}(\bk,t_0)=\frac{1}{12\,H_0^2}\int \frac{d\bp}{(2\pi)^{3/2}}\, |\bk-\bp|\,p\,{h}_{ab}(\bk-\bp,\,t_0){h}_{ab}(\bp,\,t_0)$ for the gravitational wave energy density, and defining $
	\Omega= 12\, H_0^2\,\Omega_{GW}$, the correlator becomes
\begin{align}
	{\cal C}_{\Omega \Omega}(\bk)&=\frac{1}{\Omega^2}\frac{k^3}{2\pi^2} \int \frac{d\bp_1 \,d\bp_2}{(2\pi)^3}\, |\bk-\bp_1|\, p_1\,|\bk+\bp_2|\, p_2\nonumber \\ &\times\langle h_{ab}(\bk-\bp_1,\,t_0)\,h_{ab}(\bp_1,\,t_0)\,h_{cd}(-\bk-\bp_2,\,t_0)\,h_{cd}(\bp_2,\,t_0)\rangle'\,. 
\end{align}
The current gravitational wave amplitude is related to its primordial value, calculated at the time $t_e$ when inflation ends, through the transfer function: $h_{ab}(\bk,\,t_0)= T(k)\, h_{ab}(\bk,\,t_e)$. For simplicity, from now on we will write $h_{ab}(\bk,\,t_e)$ simply as $h_{ab}(\bk)$, with the understanding that it refers to the value the tensor mode takes at the end of inflation. If we further define $\hat{T}(k)= k\,T(k)$, we can eventually express the correlator as
\begin{align}\label{Simplified_correlator}
	{\cal C}_{\Omega \Omega}(\bk)&=\frac{1}{\Omega^2}\frac{k^3}{2\pi^2} \int \frac{d\bp_1 \,d\bp_2}{(2\pi)^3}\, \hat{T}(k_1)\, \hat{T}(k_2)\,\hat{T}(k_3)\,\hat{T}(k_4)\,\langle h_{ab}(\bk_1)\,h_{ab}(\bk_2)\,h_{cd}(\bk_3)\,h_{cd}(\bk_4)\rangle'\,,
\end{align}
with $\bk_1= \bk-\bp_1$, $\bk_2= \bp_1$, $\bk_3= -\bk-\bp_2$ and $\bk_4= \bp_2$. The integration must be performed in the regime of large momenta, i.e. $p\gg k_{eq}$, where $k_{eq}$ is the scale that reentered the horizon at matter-radiation equality~\cite{Caprini:2018mtu}, since these are the momenta to which gravitational wave detectors are sensitive. For these modes, which exited the horizon towards the end of inflation and reentered during radiation domination, the transfer function takes the form
\begin{align}\label{RadiationTransferFunction}
	\hat{T}(k)=\hat{T}_{r}= \frac{3\,H_0\,\sqrt{\Omega^0_{\rm rad}}}{4\sqrt{2}}\,.
\end{align} 
In the following, when we explicitly evaluate the integrals in the large-momentum regime, we will denote the corresponding correlator with the subscript \textit{l.m.}. Finally, since we are interested in large scales, the momentum $k$ at which we evaluate the correlator is very small compared to the momenta over which we integrate, and is of the order of the scalar large-scale perturbation scale, i.e. $k \sim k_{\rm CMB}$.

	\section{Sourced correlator}
	\label{sec: Sourced correlator}
For the normalized sourced correlator eq.~\eqref{eq:CorrelatorgenericDefinition} takes the form
	\begin{align}\label{eq:CorrelatorDefinition}
		{\cal C}_{\Omega \Omega}^{\rm S}(\bk)=\frac{1}{\Omega_{GW,\rm S}^{\,2}}\frac{k^3}{2\pi^2}\int\,d\by\, e^{-i\bk\by}\,\langle \Omega_{GW,\rm S}(\bx+\by,\,t_0)\,\Omega_{GW,\rm S}(\bx,\,t_0)\rangle\,,
	\end{align}
with the fractional energy at interferometer scales for the sourced component being \begin{align}\label{omegaGW}
		\Omega_{GW,\rm S}\simeq\frac{\Omega^0_{\rm rad}}{24}\,{\cal P}_{h,\mathrm S}(k_{\rm INT})\hspace{0.5cm} \text{with} \hspace{0.5cm}  {\cal P}_{h,\mathrm S}(k_{\rm INT})=8.7\times10^{-8}\,\frac{H^4}{M_P^4}\,\frac{e^{4\pi\xi_{\rm INT}}}{\xi_{\rm INT}^6},
	\end{align}
and \begin{align}\label{omega}
		\Omega_{\rm S}= 12\, H_0^2\,\Omega_{GW,\rm S}\,.\end{align} 
		Expression~\eqref{Simplified_correlator} for the sourced correlator takes the form
		\begin{align}\label{correlator_sourced}
		{\cal C}_{\Omega \Omega}^{\rm S}	=\frac{1}{\Omega_{\rm S}^2}\frac{k^3}{2\pi^2} \int \frac{d\bp_1 \,d\bp_2}{(2\pi)^3}\, \hat{T}(k_1)\, \hat{T}(k_2)\,\hat{T}(k_3)\,\hat{T}(k_4)\,\langle h_{ab,{\rm S}}(\bk_1)\,h_{ab,{\rm S}}(\bk_2)\,h_{cd,{\rm S}}(\bk_3)\,h_{cd,{\rm S}}(\bk_4)\rangle'\,,
	\end{align}
	with $\bk_1= \bk-\bp_1$, $\bk_2= \bp_1$, $\bk_3= -\bk-\bp_2$ and $\bk_4= \bp_2$. By substituting the decomposition of the sourced tensor modes in homogeneous part and fluctuations defined in \eqref{hequalhzeroanddeltah}-\eqref{deltah} into the four-point function present in~\eqref{correlator_sourced} we obtain 
	\begin{align}\label{4pointdecomposition}
		&\langle h_{ab,{\rm S}}(\bk_1)\,h_{ab,{\rm S}}(\bk_2)\,h_{cd,{\rm S}}(\bk_3)\,h_{cd,{\rm S}}(\bk_4)\rangle= \langle h^0_{ab,{\rm S}}(\bk_1)\,h^0_{ab,{\rm S}}(\bk_2)\,h^0_{cd,{\rm S}}(\bk_3)\,h^0_{cd,{\rm S}}(\bk_4)\rangle \nonumber  \\
		+\,&\langle h^0_{ab,{\rm S}}(\bk_1)\,h^0_{ab,{\rm S}}(\bk_2)\,\delta h_{cd,{\rm S}}(\bk_3)\,\delta h_{cd,{\rm S}}(\bk_4)\rangle  + \langle\delta h_{ab,{\rm S}}(\bk_1)\,\delta h_{ab,{\rm S}}(\bk_2)\, h^0_{cd,{\rm S}}(\bk_3)\, h^0_{cd,{\rm S}}(\bk_4)\rangle \nonumber \\ + \, &4 \langle h^0_{ab,{\rm S}}(\bk_1)\,\delta h_{ab,{\rm S}}(\bk_2)\, h^0_{cd,{\rm S}}(\bk_3)\, \delta h_{cd,{\rm S}}(\bk_4)\rangle   + \langle\delta h_{ab,{\rm S}}(\bk_1)\,\delta h_{ab,{\rm S}}(\bk_2)\, \delta h_{cd,{\rm S}}(\bk_3)\, \delta h_{cd,{\rm S}}(\bk_4)\rangle \,.
	\end{align}
Plugging eq.~\eqref{4pointdecomposition} back into the correlator \eqref{correlator_sourced}, we identify three contributions: the intrinsic correlator, ${\cal C}_{\Omega \Omega}^{\rm I} $, which includes the first term in the r.h.s. of eq.~\eqref{4pointdecomposition} and involves only the homogeneous components; the extrinsic correlator, ${\cal C}_{\Omega \Omega}^{\rm E}$, which includes the sum of the next three terms in the r.h.s. of eq.~\eqref{4pointdecomposition}, containing both homogeneous components and fluctuations; and the correlator of the fluctuations, ${\cal C}_{\Omega \Omega}^{\rm F}$, arising from the fifth term in the r.h.s. of eq.~\eqref{4pointdecomposition}. The total sourced correlator of eq.~\eqref{correlator_sourced} is therefore decomposed as
	\begin{align}
		{\cal C}_{\Omega \Omega}^{\rm S} = {\cal C}_{\Omega \Omega}^{\rm I} + {\cal C}_{\Omega \Omega}^{\rm E}  + {\cal C}_{\Omega \Omega}^{\rm F}.
	\end{align}	
To proceed, we substitute the Fourier transform of eq.~\eqref{deltah}
\begin{align}
	\delta h_{ij,{\rm S}}(\bp)= -2\pi\frac{d\xi}{d\phi_0}\frac{\dot{\phi}_0}{H}\int \frac{d\bq}{(2\pi)^{3/2}}\,h_{ij,{\rm S}}(\bp-\bq)\, \zeta_{\rm V}(\bq)\,,
\end{align}
which is valid, strictly speaking, when $\bp \gg \bq$. This condition is generally satisfied, as scalar fluctuations are evaluated at CMB scales, which are much larger than the interferometer scales at which gravitational waves are measured. Then, using eq.~\eqref{two_point_function_curvature_perturbation} for the vacuum curvature perturbations we have
	\begin{align}\label{homogeneous_correlator}
		{\cal C}_{\Omega \Omega}^{\rm I}&=\frac{k^3}{2\pi^2\Omega_{\rm S}^2} \int \frac{d\bp_1 \,d\bp_2}{(2\pi)^3}\, \hat{T}(k_1)\, \hat{T}(k_2)\,\hat{T}(k_3)\,\hat{T}(k_4)\langle h^0_{ab,{\rm S}}(\bk_1)\,h^0_{ab,{\rm S}}(\bk_2)\,h^0_{cd,{\rm S}}(\bk_3)\,h^0_{cd,{\rm S}}(\bk_4)\rangle'\,,\displaybreak\\[1mm]
	\label{mixed_correlator}
		{\cal C}_{\Omega \Omega}^{\rm E}& =\frac{k^3}{2\pi^2\Omega_{\rm S}^2}\left(2\pi\frac{d\xi}{d\phi_0}\frac{\dot{\phi}_0}{H}\right)^2\int \frac{d\bp_1 \,d\bp_2\,d\bp_3}{(2\pi)^6}\, \hat{T}(k_1)\, \hat{T}(k_2)\,\hat{T}(k_3)\,\hat{T}(k_4)	 \frac{2\,\pi^2}{p_3^3}\,{\cal P}_{\zeta,{\mathrm V}} \nonumber \\
		&\times\biggl(\langle h^0_{ab,{\rm S}}(\bk_1)\,h^0_{ab,{\rm S}}(\bk_2)\,h^0_{cd,{\rm S}}(\bk_3-\bp_3)\,h^0_{cd,{\rm S}}(\bk_4+\bp_3)\rangle'  \nonumber\\ &+\,\langle h^0_{ab,{\rm S}}(\bk_1-\bp_3)\,h^0_{ab,{\rm S}}(\bk_2+\bp_3)\,h^0_{cd,{\rm S}}(\bk_3)\,h^0_{cd,{\rm S}}(\bk_4)	\rangle'\nonumber \\
		&+\,4\, \langle h^0_{ab,{\rm S}}(\bk_1)\,h^0_{ab,{\rm S}}(\bk_2-\bp_3)\,h^0_{cd,{\rm S}}(\bk_3)\,h^0_{cd,{\rm S}}(\bk_4+\bp_3)\rangle'\biggr)\,,\\[1mm]\label{fluctuations_correlator}
		{\cal C}_{\Omega \Omega}^{\rm F}& =\frac{k^3}{2\pi^2\Omega_{\rm S}^2} \left(2\pi\frac{d\xi}{d\phi_0}\frac{\dot{\phi}_0}{H}\right)^4
		\int \frac{d\bp_1 \,d\bp_2\,d\bp_3\,d\bp_4}{(2\pi)^9}\, \hat{T}(k_1)\, \hat{T}(k_2)\,\hat{T}(k_3)\,\hat{T}(k_4)
		\frac{(2\,\pi^2)^2}{p_3^3\,p_4^3}\,{\cal P}^2_{\zeta,{\mathrm V}} \nonumber \\
		&\times\biggl(\langle h^0_{ab,{\rm S}}(\bk_1-\bp_3)\,h^0_{ab,{\rm S}}(\bk_2+\bp_3)\,h^0_{cd,{\rm S}}(\bk_3-\bp_4)\,h^0_{cd,{\rm S}}(\bk_4+\bp_4)\rangle' \nonumber \\
		&+\,2\,\langle h^0_{ab,{\rm S}}(\bk_1-\bp_3)\,h^0_{ab,{\rm S}}(\bk_2-\bp_4)\,h^0_{cd,{\rm S}}(\bk_3+\bp_3)\,h^0_{cd,{\rm S}}(\bk_4+\bp_4)	\rangle'\biggr)\,,
	\end{align}
	with $\bk_1= \bk-\bp_1$, $\bk_2= \bp_1$, $\bk_3= -\bk-\bp_2$ and $\bk_4= \bp_2$.	The last correlator, which corresponds to the correlator of the fluctuations, is found to be much smaller than the other two and will therefore be neglected from now on. In order to compute the intrinsic and extrinsic sourced correlators, we use eqs.~\eqref{eq:can_norm_quantities} and~\eqref{eq:h_sourced}  to calculate the four-point function of the homogeneous components of the sourced tensor modes, denoted as $\mathbb{C}$
	\begin{align}\label{newhhhh}
		&\mathbb{C}(\bm{\kappa}_1,\bm{\kappa}_2,\bm{\kappa}_3,\bm{\kappa}_4)\nonumber\\&=\langle h^0_{ab,{\rm S}}(\bm{\kappa}_1)\,h^0_{ab,{\rm S}}(\bm{\kappa}_2)\,h^0_{cd,{\rm S}}(\bm{\kappa}_3)\,h^0_{cd,{\rm S}}(\bm{\kappa}_4)\rangle=\frac{16}{M^4_P} \langle H_{ab,{\rm S}}^0(\bm{\kappa}_1)\,H_{ab,{\rm S}}^0(\bm{\kappa}_2)\,H_{cd,{\rm S}}^0(\bm{\kappa}_3)\,H_{cd,{\rm S}}^0(\bm{\kappa}_4)\rangle \nonumber \\ & =  \frac{16}{M_P^8}\int_{-\infty}^{\tau} \frac{d\tau_1}{a(\tau_1)}\,\frac{d\tau_2}{a(\tau_2)}\,\frac{d\tau_3}{a(\tau_3)}\,\frac{d\tau_4}{a(\tau_4)} \, G_{\kappa_1}(\tau,\tau_1)\,G_{\kappa_2}(\tau,\tau_2)\,G_{\kappa_3}(\tau,\tau_3)\,G_{\kappa_4}(\tau,\tau_4) \times \mathcal{I}\,,
	\end{align}
	with $a(\tau)=-1/(H\,\tau)$, and
	\begin{align}\label{IntegralI}
		\mathcal{I}&= \int \frac{d\bq_1\,d\bq_2\,d\bq_3\,d\bq_4}{(2\pi)^6}  \nonumber \\ &\times\, e^+_a(\widehat{\bq_1})\,e^+_b(\widehat{\bm{\kappa}_1-\bq_1})\,\,e^+_a(\widehat{\bq_2})\,e^+_b(\widehat{\bm{\kappa}_2-\bq_2})\, e^+_c(\widehat{\bq_3})\,e^+_d(\widehat{\bm{\kappa}_3-\bq_3})\,e^+_c(\widehat{\bq_4})\,e^+_d(\widehat{\bm{\kappa}_4-\bq_4}) \nonumber \\ &\times\, \langle A'_+(q_1,\tau_1)\,A'_+(|\bm{\kappa}_1-\bq_1|,\tau_1)A'_+(q_2,\tau_2)\,A'_+(|\bm{\kappa}_2-\bq_2|,\tau_2)\,A'_+(q_3,\tau_3)\,A'_+(|\bm{\kappa}_3-\bq_3|,\tau_3)\nonumber \\ &\times\,A'_+(q_4,\tau_4)\,A'_+(|\bm{\kappa}_4-\bq_4|,\tau_4)\rangle\,.
	\end{align}
	Expression~\eqref{IntegralI} has been simplified by neglecting the negative-helicity photons, as, according to eq.~\eqref{eq:eomAlambda}, $A_+$ is the only helicity that undergoes amplification. Using Wick's theorem to decompose the eight-point function of gauge fields appearing in the last two lines of eq.~\eqref{IntegralI}, we find that the integral $\mathcal{I}$ can be written as the sum of six distinct integrals, i.e. $\mathcal{I}= \mathcal{I}_A\,+\,\mathcal{I}_B\,+\,\mathcal{I}_C\,+\,\mathcal{I}_D\,+\,\mathcal{I}_E\,+\,\mathcal{I}_F$, the explicit expressions of which are provided in~\eqref{integrals_Appendix}. Substituting these integrals back into Eq.~\eqref{newhhhh}, we obtain the four-point function of the homogeneous components of the sourced gravitational waves, expressed as the sum of six terms
	\begin{align}\label{sum_of_correlators}
		\mathbb{C}=  \mathbb{C}_A+   \mathbb{C}_B+ \mathbb{C}_C+ \mathbb{C}_D+ \mathbb{C}_E+ \mathbb{C}_F\,,
	\end{align}
	which can be calculated using the explicit form of the gauge field~\eqref{eq:a+appr}. Starting from $\mathbb{C}_A$ we have
	\begin{align}\label{CorrelaterA}
		&\mathbb{C}_A = \frac{4\,H^4}{M_P^8}\,\int_{-\infty}^\tau d\tau_1\,d\tau_2\,d\tau_3 \,d\tau_4\,(\tau_1\,\tau_2\,\tau_3\,\tau_4)^{1/2}\, G_{\kappa_1}(\tau,\tau_1)\,G_{\kappa_2}(\tau,\tau_2)\,G_{\kappa_3}(\tau,\tau_3)\,\,G_{\kappa_4}(\tau,\tau_4)\nonumber \\
		&\times\,\int\frac{d\bq}{(2\pi)^6} (\xi_1\,\xi_2\,\xi_3\,\xi_4)^{1/2}\,
		e^{2\pi(\xi_1+\xi_2+\xi_3+\xi_4)}\, q^{1/2}\,|\bm{\kappa}_1-\bq|^{1/2}\,|\bm{\kappa}_2+\bq|^{1/2}\,|\bm{\kappa}_2+\bm{\kappa}_4+\bq|^{1/2}\nonumber\\
		& \times\,\mathcal{A}_A(\bq,\, \bm{\kappa}_1-\bq,\, \bm{\kappa}_2+\bq,\,\bm{\kappa}_2+\bm{\kappa}_4+\bq)
		\,\, e^{-2\sqrt{-2\,\xi_1\,\tau_1}(\sqrt{q}+\sqrt{|\bm{\kappa}_1-\bq|}\,)}\,e^{-2\sqrt{-2\,\xi_2\,\tau_2}(\sqrt{q}+\sqrt{|\bm{\kappa}_2+\bq|}\,)}\nonumber \\&\times\,e^{-2\sqrt{-2\,\xi_3\,\tau_3}(\sqrt{|\bm{\kappa}_1-\bq|}+\sqrt{|\bm{\kappa}_2+\bm{\kappa}_4+\bq|}\,)}\,e^{-2\sqrt{-2\,\xi_4\,\tau_4}(\sqrt{|\bm{\kappa}_2+\bq|}+\sqrt{|\bm{\kappa}_2+\bm{\kappa}_4+\bq|}\,)}\,\,\delta(\bm{\kappa}_1+\bm{\kappa}_2+\bm{\kappa}_3+\bm{\kappa}_4)\,,
	\end{align}
	where $\xi_i\equiv \xi(\tau_i)$ are the slowly growing $\xi$ parameters satisfying $\xi_i\gg 1$. $\mathcal{A}_A$, the explicit expression of which is given in~\eqref{angular_appendix}, represents the angular part arising from the product of the helicity projectors. We simplify the expression by recalling that the fields are calculated at the end of inflation, i.e., at $\tau_{e}=-1/H$. Furthermore, since we are interested in modes that are well outside the horizon at the end of inflation, we can take the limit $k/H \to 0$. The presence of exponential terms such as $e^{-2\sqrt{-2\,\xi_1\,\tau_1}(\sqrt{q}+\sqrt{|\bm{\kappa}_1-\bq|}\,)}$, with $\xi_1\gg 1$, implies that ${\kappa}_1\,\tau_1\ll 1$. In the same way ${\kappa}_2\,\tau_2\ll 1$, ${\kappa}_3\,\tau_3\ll 1$ and ${\kappa}_4\,\tau_4\ll 1$. By Taylor expanding the propagator~\eqref{eq:prop} we have $G_{{\kappa}_i}(\tau,\,\tau_i)\simeq -H\,\tau_i^2/3$.
	
	To obtain the most general result, we will not assume a specific form for the $\xi$ parameter, but instead estimate the integral under the assumption of weak time dependence. In this case, we observe that the integral is dominated by the values of $|\tau_1|$, $|\tau_2|$, $|\tau_3|$ and $|\tau_4|$ belonging to relatively narrow windows around the following expressions, respectively:  $(\sqrt{q}+\sqrt{|\bm{\kappa}_1-\bq|}\,)^{-2}$, $(\sqrt{q}+\sqrt{|\bm{\kappa}_2+\bq|}\,)^{-2}$, $(\sqrt{|\bm{\kappa}_1-\bq|}+\sqrt{|\bm{\kappa}_2+\bm{\kappa}_4+\bq|}\,)^{-2}$ and  $(\sqrt{|\bm{\kappa}_2+\bq|}+\sqrt{|\bm{\kappa}_2+\bm{\kappa}_4+\bq|}\,)^{-2}$. We can therefore approximate the $\xi$ parameters present in expression $\mathbb{C}_A$ as
	\begin{align}\label{eq:xi1234}
		&\xi^A_1=\xi(\tau_1^A\simeq - (\sqrt{q}+\sqrt{|\bm{\kappa}_1-\bq|}\,)^{-2})\,,\nonumber\\
		&\xi^A_2=\xi(\tau_2^A\simeq - (\sqrt{q}+\sqrt{|\bm{\kappa}_2+\bq|}\,)^{-2})\,,\nonumber\\
		&\xi^A_3=\xi(\tau_3^A\simeq - (\sqrt{|\bm{\kappa}_1-\bq|}+\sqrt{|\bm{\kappa}_2+\bm{\kappa}_4+\bq|}\,)^{-2})\,,\nonumber\\
		&\xi^A_4=\xi(\tau_4^A\simeq - (\sqrt{|\bm{\kappa}_2+\bq|}+\sqrt{|\bm{\kappa}_2+\bm{\kappa}_4+\bq|}\,)^{-2})\,.
	\end{align}
	Finally, using the expression
	\begin{align}
		&\int_0^\infty dx\,x^{n-1}\,e^{-a\,\sqrt{x}}= \frac{2}{a^{2n}}\Gamma(2n)\,,
	\end{align}
	we obtain 
	\begin{align}\label{CorrelaterAfinal}
		&	\mathbb{C}_A = \frac{H^8\,\Gamma(7)^4}{3^4\,2^{36}\,M_P^8}\,\int\frac{d\bq}{(2\pi)^6} \frac{e^{2\pi(\xi^A_1+\xi^A_2+\xi^A_3+\xi^A_4)}}{(\xi^A_1\,\xi^A_2\,\xi^A_3\,\xi^A_4)^{3}}\,\mathcal{A}_A(\bq,\, \bm{\kappa}_1-\bq,\, \bm{\kappa}_2+\bq,\,\bm{\kappa}_2+\bm{\kappa}_4+\bq)\nonumber\\ &\times\,q^{1/2}\,|\bm{\kappa}_1-\bq|^{1/2}\,|\bm{\kappa}_2+\bq|^{1/2}\,|\bm{\kappa}_2+\bm{\kappa}_4+\bq|^{1/2}\,(\sqrt{q}+\sqrt{|\bm{\kappa}_1-\bq|}\,)^{-7}\,(\sqrt{q}+\sqrt{|\bm{\kappa}_2+\bq|}\,)^{-7}\nonumber\\ &\times(\sqrt{|\bm{\kappa}_1-\bq|}+\sqrt{|\bm{\kappa}_2+\bm{\kappa}_4+\bq|}\,)^{-7}(\sqrt{|\bm{\kappa}_2+\bq|}+\sqrt{|\bm{\kappa}_2+\bm{\kappa}_4+\bq|}\,)^{-7}\delta(\bm{\kappa}_1+\bm{\kappa}_2+\bm{\kappa}_3+\bm{\kappa}_4)\,.
	\end{align}
	In the same way we calculate all the other contributions to the eight-point function $\mathbb{C}$ in eq.~\eqref{sum_of_correlators}, which are given in \eqref{CorrelaterBfinal}. By substituting these expressions into eqs.~\eqref{homogeneous_correlator} and \eqref{mixed_correlator}, we can now obtain the intrinsic and extrinsic correlators of the sourced gravitational wave energy densities, whose computation is described in the next two Subsections.
	
Before proceeding with the calculation, we first parameterize the weak time dependence of the monotonically increasing function $\xi(\tau)$. We define $\xi(\tau=-1/k)\equiv \xi_k$ as the value of $\xi$ when a mode with comoving wavenumber $k$ exits the horizon, $N_k$ e-foldings before the end of inflation. For large momenta, i.e. $p\gg k_{eq}$, the parameter $\xi$ is close to its maximum value, $\xi_{\rm INT} \simeq 5$, evaluated at the time when interferometer-scale modes exit the horizon. This time is typically close to $\tau_{\rm BR}$, the time when the system enters the strong backreaction regime. Usually, this happens around  $N_{\rm BR} \simeq 10$ e-foldings before the end of inflation~\cite{Domcke:2020zez}, although there are models in which it can happen earlier, i.e. $N_{\rm BR} \simeq 40$ in~\cite{Garcia-Bellido:2023ser}.

	While the behavior of $\xi$ during the regime of strong backreaction is still object of research, lattice studies seem to suggest that this quantity stabilizes and evolves relatively slowly~\cite{Figueroa:2023oxc}. For this reason, we assume that $\xi$ becomes approximately constant for $\tau > \tau_{\rm BR}$, implying $\xi_{\rm BR} \simeq \xi_{\rm INT}$. In this case, contributions to the integrals from momenta larger than $k_{\rm BR}$ are negligible and can be safely ignored. For large momenta, the $\xi$ parameters in the denominators of the integrals can therefore be approximated as constants, contributing an overall factor of $\xi_{\rm BR}^{12}$. Meanwhile, the $\xi$ appearing in the exponents are approximated as
	\begin{align}\label{eq:ansatz_xi}
	\xi(\tau)=\left\{
	\begin{array}{ll}
		\xi_{\rm {BR}}+\delta \log(\tau_{\rm BR}/\tau) &,\,\, \tau<\tau_{\rm BR}\,,\\
		\xi_{\rm {BR}}&,\,\, \tau>\tau_{\rm BR}\,,
	\end{array}
	\right.
\end{align}
	with $\tau_{BR} = -1/k_{BR}$, accounting for contributions to the integrals from lower momenta. The parameter $\delta$ depends on the specific model under consideration, and more precisely on the number of e-foldings before the end of inflation at which backreaction becomes significant. In the cases discussed above, its value lies in the range between $0.06$~\cite{Domcke:2020zez} and $0.2$~\cite{Garcia-Bellido:2023ser}. Therefore, the exponential terms $e^{2\pi\xi_i^L}$, with $L = A, B, \dots, F$ and $i = 1,2,3,4$, appearing in expressions $\mathbb{C}_A$ to $\mathbb{C}_F$, transform as
	\begin{align}\label{exponential} e^{2\pi\xi_i^L} = e^{2\pi\xi(\tau_i^L)} = e^{2\pi\xi_{\rm BR}} \left(\frac{\tau_{\rm BR}}{\tau_i^L}\right)^{2\pi\delta}\,.
	\end{align}

	\subsection{Sourced intrinsic correlator}
	\label{subsec:homogeneous}
	In order to find the intrinsic correlator of the sourced gravitational wave energy densities we start by substituting eq.~\eqref{newhhhh} with $\bm{\kappa}_1=\bk_1= \bk-\bp_1$, $ \bm{\kappa}_2=\bk_2=\bp_1$, $ \bm{\kappa}_3=\bk_3= -\bk-\bp_2$, $\bm{\kappa}_4=\bk_4 =\bp_2$ into eq.~\eqref{homogeneous_correlator}
	\begin{align}
		{\cal C}_{\Omega \Omega}^{\rm I}(\bk)&=\frac{1}{\Omega_{\rm S}^2}\frac{k^3}{2\pi^2} \int \frac{d\bp_1 \,d\bp_2}{(2\pi)^3}\, \hat{T}(k_1)\, \hat{T}(k_2)\,\hat{T}(k_3)\,\hat{T}(k_4)\,\mathbb{C}(\bk-\bp_1,\bp_1,-\bk-\bp_2, \bp_2)\,,
	\end{align}
and we expand $\mathbb{C}$ using eqs.~\eqref{sum_of_correlators}, \eqref{CorrelaterAfinal} and \eqref{CorrelaterBfinal}. Term $\mathbb{C}_D$ contains a tadpole, as $\delta(\bm{\kappa}_1 + \bm{\kappa}_2) = \delta(\bm{\kappa}_3 + \bm{\kappa}_4) = \delta(\bk)$, and can therefore be neglected. In addition, the terms $\mathbb{C}_E$ and $\mathbb{C}_F$ become identical after resolving the delta functions.
For large values of the momenta we use~\eqref{eq:ansatz_xi} and we factor out of the integrals the $\xi^{12}_{BR}$ from the denominators, the $e^{8\pi\xi_{BR}}$ from \eqref{exponential} and the constant transfer functions~\eqref{RadiationTransferFunction}. The intrinsic correlator turns out to be
	\begin{align}\label{homogeneous_large_momenta}
		\( {\cal C}_{\Omega \Omega}^{\rm I}\)_{\rm l.m.} =	\frac{k^3\,H^8\,\Gamma(7)^4\,e^{8\,\pi\,\xi_{\rm BR}}\,\hat{T}_{r}^4}{\Omega_{\rm S}^2\,\pi^2\,3^4\,2^{37}\,M_P^8\,(2\pi)^9\,\xi^{12}_{BR} }\times \mathit{I}_{\rm I} \simeq 9.8\times 10^6 \,\biggl(\frac{k}{k_{BR}}\biggr)^3\,,
	\end{align}
	where in the last equality we have used~\eqref{omegaGW},~\eqref{omega} and the integral $ \mathit{I}_{\rm I}$ explicitly evaluated in Appendix \ref{appendix_homogeneous}, eq.~\eqref{B7}. For typical values $k\sim k_{\rm CMB}$ the factor $(k/k_{BR})^3$ is of the order of $e^{-150}$, which makes the intrinsic correlator very small.
	
	\subsection{Sourced extrinsic correlator}
	\label{subsec:hom_fluctuations}
	In order to find the extrinsic correlator of the sourced gravitational wave energy densities we substitute eq.~\eqref{newhhhh} into the three four-point functions of eq.~\eqref{mixed_correlator}, obtaining
		\begin{align}\label{correlator_extrinsic_full_expression}
	&	{\cal C}_{\Omega \Omega}^{\rm E}(\bk)=\frac{k^3\,{\cal P}_{\zeta,{\mathrm V}}}{\Omega_{\rm S}^2}\left(2\pi\frac{d\xi}{d\phi_0}\frac{\dot{\phi}_0}{H}\right)^2\int \frac{d\bp_1 \,d\bp_2\,d\bp_3}{(2\pi)^6}\, \hat{T}(|\bk-\bp_1|)\, \hat{T}(p_1)\,\hat{T}(|\bk+\bp_2)\,\hat{T}(p_2)	 \,\frac{1}{p_3^3}\, \nonumber \\
		&\times\,(	\mathbb{C}(\bk-\bp_1, \bp_1, -\bk-\bp_2-\bp_3,\bp_2+\bp_3) +\mathbb{C}(\bk-\bp_1-\bp_3, \bp_1+\bp_3, -\bk-\bp_2,\bp_2)\nonumber\\&+\,\mathbb{C}(\bk-\bp_1, \bp_1-\bp_3, -\bk-\bp_2,\bp_2+\bp_3))  \nonumber \\
		&=\,  {\cal C}_{\Omega \Omega,1}^{\rm E}(\bk)+  {\cal C}_{\Omega \Omega,2}^{\rm E}(\bk)+  {\cal C}_{\Omega \Omega,3}^{\rm E}(\bk)\,,\end{align}
	where we have considered the scalar vacuum power spectrum ${\cal P}_{\zeta, {\mathrm V}}\simeq 2\times 10^{-9}$ as a constant.
	
The sourced extrinsic correlator is therefore composed by three  terms. In both ${\cal C}_{\Omega \Omega,1}^{\rm E}(\bk)$ and ${\cal C}_{\Omega \Omega,2}^{\rm E}(\bk)$, the term $\mathbb{C}_D$ corresponds to a tadpole, leading to effects similar to those found in the intrinsic correlator. These correlators turn out to be very small due to the presence of the factor $(k/k_{BR})^3$. In the third correlator ${\cal C}_{\Omega \Omega,3}^{\rm E}(\bk)$, however, $\mathbb{C}_D$ is no longer a tadpole, but instead generates a significant scale-invariant term. Consequently, ${\cal C}_{\Omega \Omega,3}^{\rm E}(\bk)$ contains, besides the small contributions similar to those in the other two cases, denoted collectively as ${\cal C}_{\Omega \Omega,3}^{\rm E}(\bk)'$, a dominant component, which we denote as ${\cal C}_{\Omega \Omega}^{\rm S.I.}(\bk)$. This component constitutes the main result of this paper and for this reason we present its calculation separately in the next Subsection.  In Subsection~\ref{Extrinsic correlator: Term I}, we show for completeness all the other small contributions.

\subsubsection{Extrinsic correlator: Scale-invariant term ${\cal C}_{\Omega \Omega}^{\rm S.I.}(\bk)$}
\label{Extrinsic correlator: Scale-invariant term}
We now focus on calculating the scale-invariant contribution ${\cal C}_{\Omega \Omega}^{\rm S.I.}(\bk)$, which comes from term $\mathbb{C}_D$ of ${\cal C}_{\Omega \Omega,3}^{\rm E}$ with $\bm{\kappa}_1 = \mathbf{k} - \mathbf{p}_1$, $\bm{\kappa}_2 = \mathbf{p}_1 - \mathbf{p}_3$, $\bm{\kappa}_3 = -\mathbf{k} - \mathbf{p}_2$, and $\bm{\kappa}_4 = \mathbf{p}_2 + \mathbf{p}_3$.
This term is particularly important because it contains the quantity $\delta(\mathbf{k} - \mathbf{p}_3)$, which cancels the $k^3$ element in the prefactor, giving rise to a significantly large contribution.  More specifically, 
\begin{align}\label{scale_invariant_correlator}
	&{\cal C}_{\Omega \Omega}^{\rm S.I}(\bk)= \frac{H^8\,\Gamma(7)^4}{\Omega_{\rm S}^2\,3^4\,2^{34}\,M_P^8\,(2\pi)^{12}}\,\left(2\pi\frac{d\xi}{d\phi_0}\frac{\dot{\phi}_0}{H}\right)^2\, {\cal P}_{\zeta,{\mathrm V}}\int d\bp_1 \,d\bp_2\,d\bq_1 \, d\bq_2\nonumber\\&\times\,\hat{T}(|\bk-\bp_1|)\,\hat{T}(p_1)\,\hat{T}(|\bk+\bp_2|)\, \hat{T}(p_2)\,\mathcal{A}_D(\bq_1, \bk-\bp_1-\bq_1,\bq_2,-\bk-\bp_2-\bq_2)\nonumber\\
	&\times\,	\frac{q_1^{1/2}\,|\bk-\bp_1-\bq_1|^{1/2}\,q_2^{1/2}\,|\bk+\bp_2+\bq_2|^{1/2}\,	e^{2\pi(\xi_1^{\rm S.I.}+\xi_2^{\rm S.I.}+\xi_3^{\rm S.I.}+\xi_4^{\rm S.I.})}}{{(\sqrt{q_1}+\sqrt{|\bk-\bp_1-\bq_1|}\,)^{14}\,(\sqrt{q_2}+\sqrt{|\bk+\bp_2+\bq_2|}\,)^{14}}(\xi_1^{\rm S.I.}\,\xi_2^{\rm S.I.}\,\xi_3^{\rm S.I.}\,\xi_4^{\rm S.I.})^3}\,,
\end{align}
with 
\begin{align}\label{eq:xisi}
	&\xi_1^{\rm S.I.}=\xi_2^{\rm S.I.}=\xi(\tau_1^{\rm S.I.}\simeq - (\sqrt{q_1}+\sqrt{|\bk-\bp_1-\bq_1|}\,)^{-2})\,,\nonumber\\
	&\xi_3^{\rm S.I.}=\xi_4^{\rm S.I.}=\xi(\tau_3^{\rm S.I.}\simeq - (\sqrt{q_2}+\sqrt{|\bk+\bp_2+\bq_2|}\,)^{-2})\,.
\end{align}
The angular part is
\begin{align}\label{angular_explicit_si}
	\mathcal{A}_D(\bp,\bq, \br,\bs)= \frac{1}{16} (5+2\,\widehat{\bp}\,\widehat{\bq}+(\widehat{\bp}\,\widehat{\bq})^2)(5+2\,\widehat{\br}\,\widehat{\bs}+(\widehat{\br}\,\widehat{\bs})^2)\,,
\end{align}
which is calculated using 
	\begin{equation}\label{projectors}
	e^{\,\lambda}_{i}(\widehat\bk)\,e^{\,\lambda}_{j}(-\widehat\bk)= \frac{1}{2}	\, (\delta_{ij}-\widehat{k}_i\,\widehat{k}_j-i\,\lambda\,\epsilon_{ijk}\,\widehat{k}_k)\,.\end{equation}

For large momenta, we use the parametrization~\eqref{eq:ansatz_xi} for the $\xi$ functions in the exponents, while in the denominators we approximate them as simply $\xi_{BR}$. The transfer function takes the form~\eqref{RadiationTransferFunction}, and we simplify the integral by neglecting the contribution of the small $k$, wherever it appears. The scale-invariant correlator then takes the form	\begin{align}\label{scale-invariant_large_momenta}
	&\({\cal C}_{\Omega \Omega}^{\rm S.I.}(\bk)\)_{l.m.}=	\frac{H^8\,\Gamma(7)^4\,e^{8\,\pi\,\xi_{\rm BR}}\,\hat{T}_{r}^4}{\Omega_{\rm S}^2\,3^4\,2^{34}\,M_P^8\,(2\pi)^{12}\,\xi^{12}_{BR}}\, \left(2\pi\frac{d\xi}{d\phi_0}\frac{\dot{\phi}_0}{H}\right)^2\, {\cal P}_{\zeta,{\mathrm V}}\,\times \mathit{I}_{\rm S.I.}\,,\end{align}
with the integral $\mathit{I}_{\rm S.I.}$ given in \eqref{resultsi}.
Using ${\cal P}_{\zeta,{\mathrm V}}\simeq 2\times 10^{-9}$ and equations~\eqref{RadiationTransferFunction},~\eqref{omegaGW} and ~\eqref{omega},  we eventually obtain the correlator
\begin{align}
	&\({\cal C}_{\Omega \Omega}^{\rm S.I.}(\bk)\)_{l.m.} \simeq \,\frac{9.8\times 10^{-5}}{\delta^2}\left(2\pi\frac{d\xi}{d\phi_0}\frac{\dot{\phi}_0}{H}\right)^2.
\end{align}
Considering \eqref{factor} and the fact that the parameter $\delta$ takes values in the interval $0.06 \div 0.2$, the sourced scale-invariant extrinsic correlator is found to lie within the range\begin{align}
	\({\cal C}_{\Omega \Omega}^{\rm S.I.}(\bk)\)_{l.m.}\simeq 2.4\times10^{-5}   \div 2.4\times10^{-1}\,.
	\end{align}
This result will constitute the only relevant component of the sourced correlator, as it is many orders of magnitude larger than the intrinsic correlator, studied in Subsection~\ref{subsec:homogeneous}, and all other contributions to the extrinsic correlator, which we present for completeness in the next Subsection. 
	\subsubsection{Extrinsic correlator: Terms $ {\cal C}_{\Omega \Omega,1}^{\rm E}(\bk)$, ${\cal C}_{\Omega \Omega,2}^{\rm E}(\bk)$, ${\cal C}_{\Omega \Omega,3}^{\rm E}(\bk)'$}
	\label{Extrinsic correlator: Term I}
In order to find all the other terms contributing to the extrinsic correlator, which we anticipated to be very small and unobservable, we start by expanding the terms $\mathbb{C}$ in~\eqref{correlator_extrinsic_full_expression}
 using eqs.~\eqref{sum_of_correlators}, \eqref{CorrelaterAfinal} and \eqref{CorrelaterBfinal}. Using again the transfer function \eqref{RadiationTransferFunction} and the parametrization \eqref{eq:ansatz_xi} we find
	\begin{align}\label{mixed_I_large_momenta}
		\( {\cal C}_{\Omega \Omega,1}^{\rm E}(\bk)\)_{\rm l.m.}&=	\frac{k^3\,H^8\,\Gamma(7)^4\,e^{8\,\pi\,\xi_{\rm BR}}\,\hat{T}_{r}^4\,{\cal P}_{\zeta,{\mathrm V}}}{\Omega_{\rm S}^2\,3^4\,2^{36}\,M_P^8\,(2\pi)^{12}\,\xi^{12}_{BR}} \left(2\pi\frac{d\xi}{d\phi_0}\frac{\dot{\phi}_0}{H}\right)^2 \times \mathit{I}_{\rm E,1}\simeq 5.5\times10^{-1}\,\biggl(\frac{k}{k_{BR}}\biggr)^3\,,\\[1mm]
\label{mixed_II_large_momenta}
		\( {\cal C}_{\Omega \Omega,2}^{\rm E}(\bk)\)_{\rm l.m.}&=	\frac{k^3\,H^8\,\Gamma(7)^4\,e^{8\,\pi\,\xi_{\rm BR}}\,\hat{T}_{r}^4\,{\cal P}_{\zeta,{\mathrm V}}}{\Omega_{\rm S}^2\,3^4\,2^{36}\,M_P^8\,(2\pi)^{12}\,\xi^{12}_{BR}} \left(2\pi\frac{d\xi}{d\phi_0}\frac{\dot{\phi}_0}{H}\right)^2\times \mathit{I}_{\rm E,2} \simeq 4.9\times10^{-1}\,\biggl(\frac{k}{k_{BR}}\biggr)^3\,,
\\[1mm]\label{mixed_III_large_momenta}
		\({\cal C}_{\Omega \Omega,3}^{\rm E}(\bk)'\)_{\rm l.m.}&=	\frac{k^3\,H^8\,\Gamma(7)^4\,e^{8\,\pi\,\xi_{\rm BR}}\,\hat{T}_{r}^4\,{\cal P}_{\zeta,{\mathrm V}}}{\Omega_{\rm S}^2\,3^4\,2^{34}\,M_P^8\,(2\pi)^{12}\,\xi^{12}_{BR}} \left(2\pi\frac{d\xi}{d\phi_0}\frac{\dot{\phi}_0}{H}\right)^2\times \mathit{I}_{\rm E,3}\simeq 1.7\biggl(\frac{k}{k_{BR}}\biggr)^3\,,
	\end{align}
where in the final expressions we have used~\eqref{factor},~\eqref{omegaGW},~\eqref{omega} and the integrals $\mathit{I}_{\rm E,1}$, $\mathit{I}_{\rm E,2}$ and $\mathit{I}_{\rm E,3}$ evaluated, respectively, in~\eqref{result1},~\eqref{result2},~\eqref{result3}. These correlators are all very small because of the presence of $(k/k_{BR})^3$, as in the case of the intrinsic correlator.

		\section{Vacuum correlator}
	\label{sec: Vacuum correlator}
	For the normalized vacuum correlator eq.~\eqref{eq:CorrelatorgenericDefinition} becomes
	\begin{align}\label{eq:VacuumCorrelatorDefinition}
		{\cal C}_{\Omega \Omega}^{\rm V}(\bk)&=\frac{1}{\Omega_{GW,\rm V}^{\,2}}\frac{k^3}{2\pi^2}\int\,d\by\, e^{-i\bk\by}\,\langle \Omega_{GW,\rm V}(\bx+\by,\,t_0)\,\Omega_{GW,\rm V}(\bx,\,t_0)\rangle\,,
	\end{align}
	with the fractional energy at interferometer scales for the vacuum component being 
	\begin{align}\label{omegaGWvacuum}
		\Omega_{GW,\rm V}\simeq\frac{\Omega^0_{\rm rad}}{24}\,{\cal P}_{h,\mathrm V}(k_{\rm INT})\,, \hspace{0.55cm} \text{with} \hspace{0.6cm} {\cal P}_{h,\mathrm V}(k_{\rm INT})= \frac{2\,H^2}{\pi^2\,M_P^2}\,.
	\end{align}
	Defining 
	\begin{align}\label{omegaV}
		\Omega_{\rm V}= 12\, H_0^2\,\Omega_{GW,\rm V\,,}\end{align} 
	expression~\eqref{Simplified_correlator} for the vacuum correlator takes the form
	\begin{align}\label{correlator_vacuum}
		{\cal C}_{\Omega \Omega}^{\rm V}	=\frac{1}{\Omega_{\rm V}^2}\frac{k^3}{2\pi^2} \int \frac{d\bp_1 \,d\bp_2}{(2\pi)^3}\, \hat{T}(k_1)\, \hat{T}(k_2)\,\hat{T}(k_3)\,\hat{T}(k_4)\,\langle h_{ab,{\rm V}}(\bk_1)\,h_{ab,{\rm V}}(\bk_2)\,h_{cd,{\rm V}}(\bk_3)\,h_{cd,{\rm V}}(\bk_4)\rangle'\,,
	\end{align}
	with $\bk_1= \bk-\bp_1$, $\bk_2= \bp_1$, $\bk_3= -\bk-\bp_2$ and $\bk_4= \bp_2$ and  $\hat{T}(k)= k\,T(k)$. 
	The four-point function in equation~\eqref{correlator_vacuum} is decomposed using Wick's theorem, and, up to a tadpole term, is
	\begin{align}
		&	\langle h_{ab,{\rm V}}(\bk_1)\,h_{ab,{\rm V}}(\bk_2)\,h_{cd,{\rm V}}(\bk_3)\,h_{cd,{\rm V}}(\bk_4)\rangle= 2\,\langle h_{ab,{\rm V}}(\bk_1)\,h_{cd,{\rm V}}(\bk_3)\rangle\,\langle h_{ab,{\rm V}}(\bk_2)\,\,h_{cd,{\rm V}}(\bk_4)\rangle\,.
	\end{align}
	Using
	\begin{align}
		&	\langle h_{ab,{\rm V}}(\bk_1)\,h_{cd,{\rm V}}(\bk_3)\rangle =\sum_{\lambda=\pm} e^{\lambda}_{ab} (\widehat{\bk}_1)\,e^{\lambda}_{cd} (-\widehat{\bk}_1) \,\delta (\bk_1+\bk_3) \, |h_{\rm V}^\lambda(k_1)|^2\,,
	\end{align}
	and the definition $|h_{\rm V}^\lambda(k)|^2 = \frac{2\pi^2}{k^3}\,{\cal P}_{h,{\rm V}}^\lambda$, with ${\cal P}_{h,{\rm V}}^\pm  = \frac{H^2}{\pi^2\,M_P^2}$, we eventually have
	\begin{align}\label{correlator_4.7}
		{\cal C}_{\Omega \Omega}^{\rm V}(\bk)&=\frac{k^3\,H^4}{2\pi^5\,M_P^4\,\Omega_{\rm V}^2} \int d\bp\, \hat{T}(|\bk-\bp|)^2\,\hat{T}(p)^2 \frac{1}{|\bk-\bp|^3\,p^3}\,\times\,\mathcal{A}(\bk-\bp,\bp)\,.
	\end{align}
	The angular part, calculated using~\eqref{projectors}, is
	\begin{align}
		\mathcal{A}(\bk-\bp,\bp)&= \sum_{\lambda,\sigma} e^\lambda_{ab} (\widehat{\bk-\bp})\,e_{cd}^{\lambda} (-\widehat{(\bk-\bp)})\,e^\sigma_{ab} (\widehat{\bp})\,e_{cd}^{\sigma} (-\widehat{\bp})\nonumber \\ &	= \frac{1}{4} \biggl(1+6(\widehat{(\bk-\bp)}\widehat{\bp})^2+(\widehat{(\bk-\bp)}\widehat{\bp})^4\biggl)\,.
	\end{align}

The integral in~\eqref{correlator_4.7} must be evaluated over the sensitivity range of gravitational wave detectors, i.e., between momenta $p_{min}$ and $p_{max}$ that correspond to the limits of the momentum interval measurable by a given detector. As previously discussed, these momenta are much larger than the value of $k\sim k_{\rm CMB}$ under consideration, which can therefore be neglected. With this in mind, and using~\eqref{RadiationTransferFunction} and~\eqref{omegaGWvacuum}-\eqref{omegaV}, the correlator takes the form
\begin{align}
		\({\cal C}_{\Omega \Omega}^{\rm V}\)_{\rm l.m.}(\bk)\simeq \frac{k^3\,3^4\,24^2}{2\pi\,4^6\,144}\int_{p_{min}}^{p_{max}} dp\, p^2 \int d\Omega\, \frac{1}{p^6}\,	\mathcal{A}(-\bp,\bp)\,.
	\end{align}
After performing the integrals $\int d\Omega\,\mathcal{A}(-\bp,\bp)= 8\pi$ and $\int_{p_{min}}^{p_{max}} dp \,\frac{1}{p^4}\simeq \frac{1}{3\,p_{min}^3}$, the correlator eventually becomes
	\begin{align}
		\({\cal C}_{\Omega \Omega}^{\rm V}\)_{\rm l.m.}= 0.1\times \biggl(\frac{k}{p_{min}}\biggl)^3.
	\end{align}
This result, similarly to the intrinsic sourced correlator and the subdominant contributions of the extrinsic sourced correlator, is subject to a strong suppression due to the factor $(k/p_{min})^3$, and is therefore very small and unobservable for any gravitational wave detector.

	\section{Discussion and conclusions}
	\label{sec:conclusion}%
The recent evidence of a stochastic gravitational wave background reported by PTA measurements has opened a promising new observational window in modern cosmology. Such a background can arise either from the combined signals of unresolved late-time astrophysical sources, such as supermassive binary black hole mergers, or from a cosmological origin. In particular, a cosmological SGWB can reveal new details about the very early Universe in ways that previous observations could not, as the gravitational waves that make it up are produced before photon decoupling and propagate almost freely throughout the Universe after their generation.

To extract information from the cosmological gravitational wave background, we must distinguish it from its astrophysical counterpart. One approach is to analyze their anisotropies and, in particular, the correlation of these anisotropies with CMB anisotropies, which is expected to differ between the two backgrounds. In this context, the authors of~\cite{Corba_2024} computed the correlator between the curvature perturbation and the energy density of gravitational waves within the axion inflation model.

In this paper we studied the amplitude of gravitational wave anisotropies in the axion inflation model, by computing the correlator $\langle\Omega_{\rm GW}(\bx)\Omega_{\rm GW}(\by)\rangle$, which provides a measure of the observability of the correlation between scalar and tensor fluctuations. In axion inflation, the coupling of the inflaton to gauge fields implies that fluctuations arise both from the vacuum, through the standard amplification process, and from gauge fields via an inverse decay process. As a result, the correlator consists of two contributions: the correlation of the vacuum gravitational waves and the correlation of the sourced gravitational waves. Moreover, since the sourced gravitational waves consist of one part that depends only on the zero mode of the inflaton and another that depends on its fluctuations, the sourced correlator can be further decomposed into three distinct contributions: the intrinsic part, the extrinsic part, and a part containing only fluctuations.

Our analysis shows that the only relevant contribution to the correlator arises from the scale-invariant part of the extrinsic sourced component, while all other terms are negligible. For typical parameter values, the normalized sourced correlator is found to lie in the range $\mathcal{O}(10^{-5}-10^{-1})$ and in particular, it can reach values as large as $2.4\times 10^{-1}$. According to~\cite{LISACosmologyWorkingGroup:2022kbp,Mentasti:2023icu,Cui:2023dlo}, anisotropies must be relatively large to be detectable within a reasonable time frame. Our study shows that axion inflation can indeed produce observable anisotropies. Combined with the increased sensitivity of future gravitational wave detectors, this result motivates further study of angular correlations in the GW background as well as cross-correlations with CMB anisotropies in axion inflation models.

\acknowledgments I thank Lorenzo Sorbo for very useful discussions.  This work is partially supported by the US-NSF grants PHY-2112800 and PHY-2412570.

	\appendix
	\section{Sourced correlator: Full expressions}
	\label{Sourced Correlator}
	
	The integrals that compose $\mathcal{I}= \mathcal{I}_A\,+\,\mathcal{I}_B\,+\,\mathcal{I}_C\,+\,\mathcal{I}_D\,+\,\mathcal{I}_E\,+\,\mathcal{I}_F$ in  eq.~\eqref{IntegralI} are given by the expressions
	\begin{align}\label{integrals_Appendix}
		&	\mathcal{I}_A = \int \frac{d\bq}{(2\pi)^6}\, A'_+(q,\tau_1)\,A'_+(|\bm{\kappa}_1-\bq|,\tau_1)A'_+(q,\tau_2)\,A'_+(|\bm{\kappa}_2+\bq|,\tau_2)\,A'_+(|\bm{\kappa}_1-\bq|,\tau_3)\nonumber \\ &\times \,A'_+(|\bm{\kappa}_2+\bm{\kappa}_4+\bq|,\tau_3)\,A'_+(|\bm{\kappa}_2+\bq|,\tau_4)\,A'_+(|\bm{\kappa}_2+\bm{\kappa}_4+\bq|,\tau_4) \nonumber \\ &\times \mathcal{A}_A(\bq,\, \bm{\kappa}_1-\bq,\,\bm{\kappa}_2+\bq,\,\bm{\kappa}_2+\bm{\kappa}_4+\bq)\,\delta(\bm{\kappa}_1+\bm{\kappa}_2+\bm{\kappa}_3+\bm{\kappa}_4)\,,\nonumber\\[1em]
		&	\mathcal{I}_B = \int \frac{d\bq}{(2\pi)^6}\, A'_+(q,\tau_1)\,A'_+(|\bm{\kappa}_1-\bq|,\tau_1)A'_+(q,\tau_2)\,A'_+(|\bm{\kappa}_2+\bq|,\tau_2)\,A'_+(|\bm{\kappa}_2+\bq|,\tau_3)\nonumber \\ &\times \,A'_+(|\bm{\kappa}_2+\bm{\kappa}_3+\bq|,\tau_3)\,A'_+(|\bm{\kappa}_1-\bq|,\tau_4)\,A'_+(|\bm{\kappa}_2+\bm{\kappa}_3+\bq|,\tau_4) \nonumber \\ 
		&\times\,\mathcal{A}_B(\bq,\, \bm{\kappa}_1-\bq,\, \bm{\kappa}_2+\bq,\,\bm{\kappa}_2+\bm{\kappa}_3+\bq)\,\delta(\bm{\kappa}_1+\bm{\kappa}_2+\bm{\kappa}_3+\bm{\kappa}_4)\,,\nonumber\\[1em]
		&	\mathcal{I}_C = \int \frac{d\bq}{(2\pi)^6}\, A'_+(q,\tau_1)\,A'_+(|\bm{\kappa}_1-\bq|,\tau_1)A'_+(|\bm{\kappa}_3+\bq|,\tau_2)\,A'_+(|\bm{\kappa}_2+\bm{\kappa}_3+\bq|,\tau_2)\,A'_+(q,\tau_3)\nonumber \\ &\times \,A'_+(|\bm{\kappa}_3+\bq|,\tau_3)\,A'_+(|\bm{\kappa}_1-\bq|,\tau_4)\,A'_+(|\bm{\kappa}_2+\bm{\kappa}_3+\bq|,\tau_4) \nonumber \\
		&\times\,\mathcal{A}_C(\bq,\, \bm{\kappa}_1-\bq,\, \bm{\kappa}_3+\bq,\,\bm{\kappa}_2+\bm{\kappa}_3+\bq)\,\delta(\bm{\kappa}_1+\bm{\kappa}_2+\bm{\kappa}_3+\bm{\kappa}_4)\,,\nonumber\\[1em]
		&	\mathcal{I}_D = \int \frac{d\bq_1\,d\bq_2}{(2\pi)^6}\, A'_+(q_1,\tau_1)\,A'_+(|\bm{\kappa}_1-\bq_1|,\tau_1)\,A'_+(q_1,\tau_2)\,A'_+(|\bm{\kappa}_1-\bq_1|,\tau_2)\,A'_+(q_2,\tau_3)\nonumber \\ &\times \,A'_+(|\bm{\kappa}_3-\bq_2|,\tau_3)\,A'_+(q_2,\tau_4)\,A'_+(|\bm{\kappa}_3-\bq_2|,\tau_4) \nonumber \\ 
		&\times\, \mathcal{A}_D(\bq_1,\, \bm{\kappa}_1-\bq_1,\, \bq_2,\,\bm{\kappa}_3-\bq_2)\,\delta(\bm{\kappa}_1+\bm{\kappa}_2)\,\delta(\bm{\kappa}_3+\bm{\kappa}_4)\,,\nonumber\\[1em]
		&	\mathcal{I}_E= \int \frac{d\bq_1\,d\bq_2}{(2\pi)^6}\, A'_+(q_1,\tau_1)\,A'_+(|\bm{\kappa}_1-\bq_1|,\tau_1)\,A'_+(q_2,\tau_2)\,A'_+(|\bm{\kappa}_2-\bq_2|,\tau_2)\,A'_+(q_1,\tau_3)\nonumber \\ &\times \,A'_+(|\bm{\kappa}_1-\bq_1|,\tau_3)\,A'_+(q_2,\tau_4)\,A'_+(|\bm{\kappa}_2-\bq_2|,\tau_4) \nonumber \\ 
		&\times\,\mathcal{A}_E(\bq_1,\, \bm{\kappa}_1-\bq_1,\, \bq_2,\,\bm{\kappa}_2-\bq_2) \,\delta(\bm{\kappa}_2+\bm{\kappa}_4)\,\delta(\bm{\kappa}_1+\bm{\kappa}_3)\,,\nonumber\\[1em]
		&	\mathcal{I}_F= \int \frac{d\bq_1\,d\bq_2}{(2\pi)^6}\, A'_+(q_1,\tau_1)\,A'_+(|\bm{\kappa}_1-\bq_1|,\tau_1)\,A'_+(q_2,\tau_2)\,A'_+(|\bm{\kappa}_2-\bq_2|,\tau_2)\,A'_+(q_2,\tau_3)\nonumber \\ &\times \,A'_+(|\bm{\kappa}_2-\bq_2|,\tau_3)\,A'_+(q_1,\tau_4)\,A'_+(|\bm{\kappa}_1-\bq_1|,\tau_4) \nonumber \\ 
		&\times\,\mathcal{A}_F(\bq_1,\, \bm{\kappa}_1-\bq_1,\, \bq_2,\,\bm{\kappa}_2-\bq_2)\,\delta(\bm{\kappa}_1+\bm{\kappa}_4)\,\delta(\bm{\kappa}_2+\bm{\kappa}_3)\,,
	\end{align}
	where we have collected the angular parts inside the functions $\mathcal{A}$: 
	\begin{align}\label{angular_appendix}
		&\mathcal{A}_A(\bp_1, \bp_2,\bp_3,\bp_4)= \mathcal{A}_B(\bp_1, \bp_2,\bp_3,\bp_4) =(((( e^+_a(\widehat{\bp_1})\,e^+_{\beta}(\widehat{-\bp_1})\,e^+_b(\widehat{\bp_2})\,e^+_c(\widehat{-\bp_2})\, e^+_\alpha(\widehat{\bp_3})\nonumber \\&\times\,e^+_\gamma(\widehat{-\bp_3})\,e^+_\delta(\widehat{\bp_4})\,e^+_d(\widehat{-\bp_4})+\,(a\leftrightarrow b))+ (\alpha\leftrightarrow \beta))+(c\leftrightarrow d))+(\gamma\leftrightarrow \delta))\delta_{a\alpha}\,\delta_{b\beta}\,\delta_{c\gamma}\,\delta_{d\delta}\,,\nonumber\\[1em]
		&\mathcal{A}_C(\bp_1, \bp_2,\bp_3,\bp_4)=(((( e^+_a(\widehat{\bp_1})\,e^+_{c}(\widehat{-\bp_1})\,e^+_b(\widehat{\bp_2})\,e^+_\gamma(\widehat{-\bp_2})\, e^+_d(\widehat{\bp_3})\,e^+_\alpha(\widehat{-\bp_3})\,e^+_\beta(\widehat{\bp_4})\,e^+_\delta(\widehat{-\bp_4})\nonumber \\&+\,(a\leftrightarrow b))+ (\alpha\leftrightarrow \beta))+(c\leftrightarrow d))+(\gamma\leftrightarrow \delta))\delta_{a\alpha}\,\delta_{b\beta}\,\delta_{c\gamma}\,\delta_{d\delta}\,,\nonumber\\[1em]
		&\mathcal{A}_D(\bp_1, \bp_2,\bp_3,\bp_4)=(( e^+_a(\widehat{\bp_1})\,e^+_{\alpha}(\widehat{-\bp_1})\,e^+_b(\widehat{\bp_2})\,e^+_\beta(\widehat{-\bp_2})\, e^+_c(\widehat{\bp_3})\,e^+_\gamma(\widehat{-\bp_3})\,e^+_d(\widehat{\bp_4})e^+_\delta(\widehat{-\bp_4})\nonumber \\&+(\alpha\leftrightarrow \beta))+ (\gamma\leftrightarrow \delta))\delta_{a\alpha}\,\delta_{b\beta}\,\delta_{c\gamma}\,\delta_{d\delta}\,,\nonumber\\[1em]
		&\mathcal{A}_E(\bp_1, \bp_2,\bp_3,\bp_4)=\mathcal{A}_F(\bp_1, \bp_2,\bp_3,\bp_4)=(( e^+_a(\widehat{\bp_1})\,e^+_{c}(\widehat{-\bp_1})\,e^+_b(\widehat{\bp_2})\,e^+_d(\widehat{-\bp_2})\, e^+_\alpha(\widehat{\bp_3})\nonumber \\&\times\,e^+_\gamma(\widehat{-\bp_3})\,e^+_\beta(\widehat{\bp_4})e^+_\delta(\widehat{-\bp_4})+(c\leftrightarrow d))+ (\gamma\leftrightarrow \delta))\delta_{a\alpha}\,\delta_{b\beta}\,\delta_{c\gamma}\,\delta_{d\delta}\,.
	\end{align}

The correlators $\mathbb{C}_B$ to $\mathbb{C}_F$ in eq.~\eqref{sum_of_correlators}, calculated in a similar way as $\mathbb{C}_A$ in \eqref{CorrelaterA}, take the form
	\begin{align}\label{CorrelaterBfinal}
	&	\mathbb{C}_B = \frac{H^8\,\Gamma(7)^4}{3^4\,2^{36}\,M_P^8}\,\int\frac{d\bq}{(2\pi)^6} \frac{e^{2\pi(\xi^B_1+\xi^B_2+\xi^B_3+\xi^B_4)}}{(\xi^B_1\,\xi^B_2\,\xi^B_3\,\xi^B_4)^{3}}\,\mathcal{A}_B(\bq, \,\bm{\kappa}_1-\bq,\, \bm{\kappa}_2+\bq,\,\bm{\kappa}_2+\bm{\kappa}_3+\bq)\nonumber\\&\times\,q^{1/2}\,|\bm{\kappa}_1-\bq|^{1/2}\,|\bm{\kappa}_2+\bq|^{1/2}\,|\bm{\kappa}_2+\bm{\kappa}_3+\bq|^{1/2}\,(\sqrt{q}+\sqrt{|\bm{\kappa}_1-\bq|}\,)^{-7}\,(\sqrt{q}+\sqrt{|\bm{\kappa}_2+\bq|}\,)^{-7}\nonumber\\ &\times(\sqrt{|\bm{\kappa}_1-\bq|}+\sqrt{|\bm{\kappa}_2+\bm{\kappa}_3+\bq|}\,)^{-7}(\sqrt{|\bm{\kappa}_2+\bq|}+\sqrt{|\bm{\kappa}_2+\bm{\kappa}_3+\bq|}\,)^{-7}\delta(\bm{\kappa}_1+\bm{\kappa}_2+\bm{\kappa}_3+\bm{\kappa}_4)\,,\nonumber\\[1em]
	&	\mathbb{C}_C = \frac{H^8\,\Gamma(7)^4}{3^4\,2^{36}\,M_P^8}\,\int\frac{d\bq}{(2\pi)^6} \frac{e^{2\pi(\xi^C_1+\xi^C_2+\xi^C_3+\xi^C_4)}}{(\xi^C_1\,\xi^C_2\,\xi^C_3\,\xi^C_4)^{3}}\,\mathcal{A}_C(\bq,\, \bm{\kappa}_1-\bq,\, \bm{\kappa}_3+\bq,\,\bm{\kappa}_2+\bm{\kappa}_3+\bq)\nonumber\\&\times\,q^{1/2}\,|\bm{\kappa}_1-\bq|^{1/2}\,|\bm{\kappa}_3+\bq|^{1/2}\,|\bm{\kappa}_2+\bm{\kappa}_3+\bq|^{1/2}\,(\sqrt{q}+\sqrt{|\bm{\kappa}_1-\bq|}\,)^{-7}\,(\sqrt{q}+\sqrt{|\bm{\kappa}_3+\bq|}\,)^{-7}\nonumber\\ &\times(\sqrt{|\bm{\kappa}_1-\bq|}+\sqrt{|\bm{\kappa}_2+\bm{\kappa}_3+\bq|}\,)^{-7}(\sqrt{|\bm{\kappa}_3+\bq|}+\sqrt{|\bm{\kappa}_2+\bm{\kappa}_3+\bq|}\,)^{-7}\delta(\bm{\kappa}_1+\bm{\kappa}_2+\bm{\kappa}_3+\bm{\kappa}_4)\,,\nonumber\\[1em]
	&	\mathbb{C}_D= \frac{H^8\,\Gamma(7)^4}{3^4\,2^{36}\,M_P^8}\,\int\frac{d\bq_1\,d\bq_2}{(2\pi)^6} \frac{e^{2\pi(\xi^D_1+\xi^D_2+\xi^D_3+\xi^D_4)}}{(\xi^D_1\,\xi^D_2\,\xi^D_3\,\xi^D_4)^{3}}\,\mathcal{A}_D(\bq_1,\, \bm{\kappa}_1-\bq_1,\, \bq_2,\,\bm{\kappa}_3-\bq_2)\nonumber\\ &\times\,q_1^{1/2}\,q_2^{1/2}\,|\bm{\kappa}_1-\bq_1|^{1/2}\,|\bm{\kappa}_3-\bq_2|^{1/2}\,(\sqrt{q_1}+\sqrt{|\bm{\kappa}_1-\bq_1|}\,)^{-14}\,(\sqrt{q_2}+\sqrt{|\bm{\kappa}_3-\bq_2|}\,)^{-14}\nonumber\\
	& \times \,\delta(\bm{\kappa}_1+\bm{\kappa}_2)\,\delta(\bm{\kappa}_3+\bm{\kappa}_4)\,,\nonumber\\[1em]
	&	\mathbb{C}_E= \frac{H^8\,\Gamma(7)^4}{3^4\,2^{36}\,M_P^8}\,\int\frac{d\bq_1\,d\bq_2}{(2\pi)^6} \frac{e^{2\pi(\xi^E_1+\xi^E_2+\xi^E_3+\xi^E_4)}}{(\xi^E_1\,\xi^E_2\,\xi^E_3\,\xi^E_4)^{3}}\,\mathcal{A}_E(\bq_1,\, \bm{\kappa}_1-\bq_1,\, \bq_2,\,\bm{\kappa}_2-\bq_2)\nonumber\\ &\times\,q_1^{1/2}\,q_2^{1/2}\,|\bm{\kappa}_1-\bq_1|^{1/2}\,|\bm{\kappa}_2-\bq_2|^{1/2}\,(\sqrt{q_1}+\sqrt{|\bm{\kappa}_1-\bq_1|}\,)^{-14}\,(\sqrt{q_2}+\sqrt{|\bm{\kappa}_2-\bq_2|}\,)^{-14}\nonumber\\
	& \times \,\delta(\bm{\kappa}_2+\bm{\kappa}_4)\,\delta(\bm{\kappa}_1+\bm{\kappa}_3)\,,\nonumber\\[1em]
	&	\mathbb{C}_F= \frac{H^8\,\Gamma(7)^4}{3^4\,2^{36}\,M_P^8}\,\int\frac{d\bq_1\,d\bq_2}{(2\pi)^6} \frac{e^{2\pi(\xi^F_1+\xi^F_2+\xi^F_3+\xi^F_4)}}{(\xi^F_1\,\xi^F_2\,\xi^F_3\,\xi^F_4)^{3}}\,\mathcal{A}_F(\bq_1,\, \bm{\kappa}_1-\bq_1,\, \bq_2,\,\bm{\kappa}_2-\bq_2)\nonumber\\ &\times\,q_1^{1/2}\,q_2^{1/2}\,|\bm{\kappa}_1-\bq_1|^{1/2}\,|\bm{\kappa}_2-\bq_2|^{1/2}\,(\sqrt{q_1}+\sqrt{|\bm{\kappa}_1-\bq_1|}\,)^{-14}\,(\sqrt{q_2}+\sqrt{|\bm{\kappa}_2-\bq_2|}\,)^{-14}\nonumber\\
	& \times \,\delta(\bm{\kappa}_1+\bm{\kappa}_4)\,\delta(\bm{\kappa}_2+\bm{\kappa}_3)\,.
\end{align}

The parameters $\xi_i^L = \xi(\tau_i^L)$ have a form similar to \eqref{eq:xi1234}, with the temporal variables evaluated, in each case, at the momenta appearing in the denominators of the respective expressions, i.e.:	\small
	\begin{equation}\label{times}
			\setlength{\tabcolsep}{20pt}
		\renewcommand{\arraystretch}{1.5}
		\begin{array}{l@{\extracolsep{15pt}}l} 
	\tau_1^A = \tau_1^B = \tau_1^C \simeq -\left( \sqrt{q} + \sqrt{|\boldsymbol{\kappa}_1 - \mathbf{q}|} \right)^{-2}, 
			& \tau_2^A = \tau_2^B \simeq -\left( \sqrt{q} + \sqrt{|\boldsymbol{\kappa}_2 + \mathbf{q}|} \right)^{-2}, \\
			\tau_3^A \simeq -\left( \sqrt{|\boldsymbol{\kappa}_1 - \mathbf{q}|} + \sqrt{|\boldsymbol{\kappa}_2 + \boldsymbol{\kappa}_4 + \mathbf{q}|} \right)^{-2}, 
			& \tau_4^B \simeq -\left( \sqrt{|\boldsymbol{\kappa}_2 + \mathbf{q}|} + \sqrt{|\boldsymbol{\kappa}_2 + \boldsymbol{\kappa}_3 + \mathbf{q}|} \right)^{-2}, \\
			\tau_4^A \simeq -\left( \sqrt{|\boldsymbol{\kappa}_2 + \mathbf{q}|} + \sqrt{|\boldsymbol{\kappa}_2 + \boldsymbol{\kappa}_4 + \mathbf{q}|} \right)^{-2}, 
			&  \tau_2^C \simeq -\left( \sqrt{q} + \sqrt{|\boldsymbol{\kappa}_3 + \mathbf{q}|} \right)^{-2}, \\
			\tau_3^B = \tau_3^C \simeq -\left( \sqrt{|\boldsymbol{\kappa}_1 - \mathbf{q}|} + \sqrt{|\boldsymbol{\kappa}_2 + \boldsymbol{\kappa}_3 + \mathbf{q}|} \right)^{-2}, 
			& \tau_4^C \simeq -\left(\sqrt{|\boldsymbol{\kappa}_3 + \mathbf{q}|} + \sqrt{|\boldsymbol{\kappa}_2 + \boldsymbol{\kappa}_3 + \mathbf{q}|} \right)^{-2}, \\
			\tau_1^E = \tau_2^E = \tau_1^F = \tau_2^F \simeq -\left( \sqrt{q_1} + \sqrt{|\boldsymbol{\kappa}_1 - \mathbf{q}_1|} \right)^{-2}, 
			& \tau_1^D = \tau_2^D \simeq -\left( \sqrt{q_1} + \sqrt{|\boldsymbol{\kappa}_1 - \mathbf{q}_1|} \right)^{-2}, \\
			\tau_3^E = \tau_4^E = \tau_3^F = \tau_4^F \simeq -\left( \sqrt{q_2} + \sqrt{|\boldsymbol{\kappa}_2 - \mathbf{q}_2|} \right)^{-2}, 
			& \tau_3^D = \tau_4^D \simeq -\left( \sqrt{q_2} + \sqrt{|\boldsymbol{\kappa}_3 - \mathbf{q}_2|} \right)^{-2}.
		\end{array}
	\end{equation}\normalsize

	\section{Sourced intrinsic correlator: Evaluation of integrals}
	\label{appendix_homogeneous}
	The integral $\mathit{I}_{\rm I}$ in eq.~\eqref{homogeneous_large_momenta} of the sourced intrinsic correlator in the regime of large momenta has the form
	\begin{align}\label{homogeneous_correlator_explicit_largemomenta}		&\mathit{I}_{\rm I}=\frac{1}{k_{BR}^{8\pi\delta}}\int d\bp_1 \,d\bp_2\,d\bq\, \nonumber\\
		&\times\,\biggl(\frac{q^{1/2}\,|\bp_1+\bq|\,|\bp_1+\bp_2+\bq|^{1/2} \,\mathcal{A}_A(\bq, -\bp_1-\bq, \bp_1+\bq,\bp_1+\bp_2+\bq)}{(\sqrt{q}+\sqrt{|\bp_1+\bq|}\,)^{14}\,(\sqrt{|\bp_1+\bq|}+\sqrt{|\bp_1+\bp_2+\bq|}\,)^{14}\,(\tau_{\rm {I},1}^A\,\tau_{\rm {I},2}^A\,\tau_{\rm {I},3}^A\,\tau_{\rm {I},4}^A)^{2\pi\delta}}\nonumber\\
		&+\,\frac{q^{1/2}\,|\bp_1+\bq|\,|\bp_1+\bq-\bp_2|^{1/2}\,\mathcal{A}_B(\bq, -\bp_1-\bq, \bp_1+\bq,\bp_1+\bq-\bp_2)}{(\sqrt{q}+\sqrt{|\bp_1+\bq|}\,)^{14}\,(\sqrt{|\bp_1+\bq|}+\sqrt{|\bp_1+\bq-\bp_2|}\,)^{14}\,(\tau_{\rm {I},1}^B\,\tau_{\rm {I},2}^B\,\tau_{\rm {I},3}^B\,\tau_{\rm {I},4}^B)^{2\pi\delta}}\nonumber \\
		&+\,\frac{q^{1/2}\,|\bp_1+\bq|^{1/2}\,|\bq-\bp_2|^{1/2}\,|\bp_1+\bq-\bp_2|^{1/2}}{(\sqrt{q}+\sqrt{|\bp_1+\bq|}\,)^7\,(\sqrt{q}+\sqrt{|\bq-\bp_2|}\,)^7\,(\sqrt{|\bq-\bp_2|}+\sqrt{|\bp_1+\bq-\bp_2|}\,)^7}\nonumber \\&\times \frac{\mathcal{A}_C(\bq, -\bp_1-\bq, \bq-\bp_2,\bp_1+\bq-\bp_2)}{(\sqrt{|\bp_1+\bq|}+\sqrt{|\bp_1+\bq-\bp_2|}\,)^7\,(\tau_{\rm {I},1}^C\,\tau_{\rm {I},2}^C\,\tau_{\rm {I},3}^C\,\tau_{\rm {I},4}^C)^{2\pi\delta}}
		\nonumber\\& + 2\,\frac{p_2^{1/2}\,q^{1/2}\,|\bp_1+\bp_2|^{1/2}\,|\bp_1-\bq|^{1/2}\,\mathcal{A}_E(\bp_2, -\bp_1-\bp_2, \bq,\bp_1-\bq)}{(\sqrt{p_2}+\sqrt{|\bp_1+\bp_2|}\,)^{14}\,(\sqrt{q}+\sqrt{|\bp_1-\bq|}\,)^{14}\,(\tau_{\rm {I},1}^E\,\tau_{\rm {I},2}^E\,\tau_{\rm {I},3}^E\,\tau_{\rm {I},4}^E)^{2\pi\delta}}\biggr)\,,
	\end{align}
	where we have used $\tau_{BR} = -1/k_{BR}$ in equation \eqref{exponential} and neglected the parameter $k$ wherever it appeared, as it is small compared to the large momenta considered in the integral. The time variables of equation \eqref{times} in this case take the form
	\begin{equation*}
	\renewcommand{\arraystretch}{1.2}
	\setlength{\tabcolsep}{6pt}
{\fontsize{9.5}{11}
	\begin{array}{l@{\extracolsep{15pt}}l} 
		\tau_{I,1}^{A} = \tau_{I,2}^{A} = \tau_{I,1}^{B} = \tau_{I,2}^{B} = \tau_{I,1}^{C} \simeq -\left( \sqrt{q} + \sqrt{|\mathbf{p}_1 + \mathbf{q}|} \right)^{-2},  &  
		\tau_{I,2}^{C} \simeq -\left( \sqrt{q} + \sqrt{|\mathbf{q} - \mathbf{p}_2|} \right)^{-2},  \\
		\tau_{I,3}^{A} = \tau_{I,4}^{A} \simeq -\left( \sqrt{|\mathbf{p}_1 + \mathbf{q}|} + \sqrt{|\mathbf{p}_1 + \mathbf{p}_2 + \mathbf{q}|} \right)^{-2},  & 	\tau_{I,3}^{C} \simeq -\left( \sqrt{|\mathbf{q} - \mathbf{p}_2|} + \sqrt{|\mathbf{p}_1 + \mathbf{q} - \mathbf{p}_2|} \right)^{-2},  \\
		\tau_{I,3}^{B} = \tau_{I,4}^{B} \simeq -\left( \sqrt{|\mathbf{p}_1 + \mathbf{q}|} + \sqrt{|\mathbf{p}_1 + \mathbf{q}  - \mathbf{p}_2|} \right)^{-2},  &  
		 \tau_{I,1}^{E} = \tau_{I,2}^{E} \simeq -\left( \sqrt{p_2} + \sqrt{|\mathbf{p}_1 + \mathbf{p}_2|} \right)^{-2},  \\
		\tau_{I,4}^{C} \simeq -\left( \sqrt{|\mathbf{p}_1 + \mathbf{q}|} + \sqrt{|\mathbf{p}_1 + \mathbf{q} - \mathbf{p}_2|} \right)^{-2},  &  \tau_{I,3}^{E} = \tau_{I,4}^{E} \simeq -\left( \sqrt{q} + \sqrt{|\mathbf{p}_1 - \mathbf{q}|} \right)^{-2}\,.
	\end{array}}
	\end{equation*}
\normalsize

To compute the integral, we consider all possible orderings of the magnitudes of $\bp_1$,  $\bp_2$ and $\bq$ and integrate over one momentum at a time, keeping only the dominant terms in each expression. For example, in the case $p_1 > p_2 > q$, the expression $(\sqrt{|\bq - \bp_2|} + \sqrt{|\bp_1 + \bq - \bp_2|})^7$ simplifies to $p_1^{7/2}$, and the integration is carried out as
	\begin{align}
		\int d\bp_1 \,d\bp_2\,d\bq= \int^{k_{BR}} dp_1 \,p_1^2 \int^{p_1} dp_2\, p_2^2  \int^{p_2} dq\, q^2 \,\int d\Omega\,,
	\end{align}
	with 
	\begin{align}\label{solid_angle}
		\int d\Omega= \prod_{i= p_1,p_2,q}\int  d\theta_i\,d\phi_i \sin(\theta_{i})\,.
	\end{align}
 A key simplification arises when the dominant momentum appears in both square roots in a sum. For instance, the expression $(\sqrt{|\bp_1 + \bq|} + \sqrt{|\bp_1 + \bq - \bp_2|})^7$, when $p_1$ is the maximum momentum, becomes $2^7\,p_1^{7/2}$. In such cases, the suppression factor of $2^7$ in the denominator makes the corresponding term negligible. Therefore, although we would theoretically need to consider all six possible orderings of the three variables, the number of relevant cases is reduced by focusing only on terms without the large $2^7$ suppression.

For the first expression of \eqref{homogeneous_correlator_explicit_largemomenta} the only ordering that does not lead to a suppression is $p_2>p_1>q$, for which the expression becomes
	\begin{align}\label{integralAlargemomenta}
		\frac{1}{k_{BR}^{8\pi\delta}}&\,\int^{k_{BR}} dp_2\, p_2^{4\pi\delta-9/2} \,\int^{p_2} dp_1\,p_1^{4\pi\delta-4} \int^{p_1} dq\,q^{5/2}\,\int d\Omega\, \mathcal{A}_A(\bq, -\bp_1, \bp_1,\bp_2)\nonumber\\
		&= \frac{1.4\times 10^4}{k_{BR}^3}\frac{2}{7\,(4\pi\delta+1/2)\,(8\pi\delta-3)},\hspace{0.4cm}\text{for} \hspace{0.2cm}\delta>\frac{3}{8\pi}\,.
	\end{align}
	The number $1.4\times 10^4$ is the result of the numerical integration of the angular part over the solid angle~\eqref{solid_angle}.	The second integral in \eqref{homogeneous_correlator_explicit_largemomenta} shares the same dependence on the momenta as the first one and therefore gives the same result. For the third integral in \eqref{homogeneous_correlator_explicit_largemomenta}, there is no ordering of the momenta that avoids the significant $2^7$ suppression, so it can be neglected. Finally, the last integral in \eqref{homogeneous_correlator_explicit_largemomenta} receives contributions only from the orderings $p_1 > q > p_2$ and $p_1 > p_2 > q$. Since these two orderings give rise to the same integral, we need to calculate only one of them:
	\begin{align}\label{integralElargemomenta}
		\frac{2}{k_{BR}^{8\pi\delta}}&\,\int^{k_{BR}} dp_1\, p_1^{8\pi\delta-11} \,\int^{p_1} dp_2\,p_2^{5/2} \int^{p_2} dq\,q^{5/2}\,\int d\Omega\, \mathcal{A}_E(\bp_2, -\bp_1, \bq,\bp_1)\nonumber\\
		&= \frac{1.8\times 10^3}{k_{BR}^3}\frac{4}{49\,(8\pi\delta-3)}\,,\hspace{0.4cm}\text{for} \hspace{0.2cm}\delta>\frac{3}{8\pi}\,.
	\end{align}
Summing all contributions, we obtain
	\begin{align}
		\mathit{I}_{\rm I}= \frac{8.1\times 10^3}{k_{BR}^3\,(4\pi\delta+1/2)\,(8\pi\delta-3)}+ \frac{2.9\times10^2}{k_{BR}^3\,(8\pi\delta-3)}\,,\hspace{0.2cm}\text{for} \hspace{0.1cm}\delta>\frac{3}{8\pi}\,.
	\end{align}

In practice, the parameter $\delta$ takes values in the range $0.06 - 0.2$. However, since this integral contributes to a correlator that turns out to be negligible for any value of $\delta$ (due to the suppression factor $(k/k_{\rm BR})^3$), we can adopt the value $\delta = 0.2$ that allows us to simplify the integrals, while still providing a reasonable estimate. For $\delta=0.2$, the integral becomes
	\begin{align}\label{B7}
		\mathit{I}_{\rm I}\simeq\frac{1.5\times 10^3}{k^3}\,\biggl(\frac{k}{k_{BR}}\biggr)^3\,,
	\end{align}
	that gives the result present in \eqref{homogeneous_large_momenta}.
	
	\section{Sourced extrinsic correlator: Evaluation of integrals}
	\label{appendix_mixedI}
	\subsection{Term: ${\cal C}_{\Omega \Omega}^{\rm S.I.}(\bk)$}
	\label{Term 3: Scale-invariant part}
	The integral $\mathit{I}_{\rm S.I.}$ in eq.~\eqref{scale-invariant_large_momenta}  of the scale-invariant part of ${\cal C}_{\Omega \Omega,3}^{\rm E}(\bk)$ in the regime of large momenta has the form
	\begin{align}
		&\mathit{I}_{\rm S.I.} =\frac{1}{k_{BR}^{8\pi\delta}}\int d\bp_1 \,d\bp_2\,d\bq_1\,d\bq_2 \nonumber\\
		&\times\,\frac{q_1^{1/2}\,|\bp_1+\bq_1|^{1/2}\,q_2^{1/2}\,|\bp_2+\bq_2|^{1/2}\,\mathcal{A}_D(\bq_1,-\bp_1-\bq_1, \bq_2,-\bp_2-\bq_2)}{(\sqrt{q_1}+\sqrt{|\bp_1+\bq_1|}\,)^{14}\,(\sqrt{q_2}+\sqrt{|\bp_2+\bq_2|}\,)^{14}\,(\tau_1^{\rm S.I.}\,\tau_2^{\rm S.I.}\,\tau_3^{\rm S.I.}\,\tau_4^{\rm S.I.})^{2\pi\delta}}\,,
	\end{align}
	with
	\begin{align}
		&\tau_1^{\rm S.I.}=\tau_2^{\rm S.I.} \simeq - (\sqrt{q_1}+\sqrt{|\bp_1+\bq_1|}\,)^{-2}\,,\nonumber\\
		&\tau_3^{\rm S.I.}=\tau_4^{\rm S.I.}\simeq - (\sqrt{q_2}+\sqrt{|\bp_2+\bq_2|}\,)^{-2}\,.\nonumber
	\end{align}
	The computation is done by integrating over one momentum at a time, accounting for all possible orderings of the four variables, as done in Appendix~\ref{appendix_homogeneous}. However, when $q_1 > p_1$ or $q_2 > p_2$, a large suppression factor of $2^{14}$ appears in each of the two parentheses in the denominator. Thus, we can restrict our calculations to the cases where $p_1 > q_1$ and $p_2 > q_2$, for which the integral becomes
	\begin{align}
	\mathit{I}_{\rm S.I.}&=\frac{1}{k_{BR}^{8\pi\delta}} \int d\bp_1\,d\bp_2\,d\bq_1\,d\bq_2 \,q_1^{1/2}\,q_2^{1/2}\,p_1^{4\pi\delta-13/2}\,p_2^{4\pi\delta-13/2}\,\mathcal{A}_D(\bq_1,-\bp_1, \bq_2,-\bp_2)\,.\end{align}
	Finally, due to the symmetry of the integral under the exchanges $\bp_1 \leftrightarrow \bp_2$ and $\bq_1 \leftrightarrow \bq_2$, it simplifies to
	\begin{align}\label{resultsi}
	\mathit{I}_{\rm S.I.}&	=\frac{1}{k_{BR}^{8\pi\delta}}\,\biggl(\int^{k_{BR}} dp\,p^{4\pi\delta-9/2} \,\int^{p}dq\,q^{5/2}\biggl)^2\,\int d\Omega \,\mathcal{A}_D(\bq_1,-\bp_1, \bq_2,-\bp_2)\nonumber\\
		&\simeq 4.4\times 10^4\,\biggl(\frac{1}{14\pi\delta}\biggl)^2 \,,
	\end{align}
	where in $d\Omega$ we have collected the angular integrations on the polar angles of all the four variables, i.e. $d\Omega=d\Omega_{p_1}\,d\Omega_{p_2}\,d\Omega_{q_1}\,d\Omega_{q_2}$.

	\subsection{Term: ${\cal C}_{\Omega \Omega,1}^{\rm E}(\bk)$}
	\label{Term 1}
	The first contribution to the extrinsic correlator of the gravitational wave energy densities in the regime of large momenta, eq.~\eqref{mixed_I_large_momenta}, is found by evaluating the integral
	\begin{align}\label{mixed_I_correlator_explicit_large_momenta}
		&\mathit{I}_{\rm E,1}=\frac{1}{k_{BR}^{8\pi\delta}}\int d\bp_1 \,d\bp_2\,d\bp_3\,d\bq\,\frac{1}{p_3^3} \nonumber\\
		&\times\,\biggl(\frac{q^{1/2}\,|\bp_1+\bq|\,|\bp_1+\bp_2+\bp_3+\bq|^{1/2} \,\mathcal{A}_A(\bq, -\bp_1-\bq, \bp_1+\bq,\bp_1+\bp_2+\bp_3+\bq)}{(\sqrt{q}+\sqrt{|\bp_1+\bq|}\,)^{14}\,(\sqrt{|\bp_1+\bq|}+\sqrt{|\bp_1+\bp_2+\bp_3+\bq|}\,)^{14}\,(\tau_{\rm {E1},1}^A\,\tau_{\rm {E1},2}^A\,\tau_{\rm {E1},3}^A\,\tau_{\rm {E1},4}^A)^{2\pi\delta}}\nonumber\\
		&+\,\frac{q^{1/2}\,|\bp_1+\bq|\,|\bp_1+\bq-\bp_2-\bp_3|^{1/2}\,\mathcal{A}_B(\bq, -\bp_1-\bq, \bp_1+\bq,\bp_1+\bq-\bp_2-\bp_3)}{(\sqrt{q}+\sqrt{|\bp_1+\bq|}\,)^{14}\,(\sqrt{|\bp_1+\bq|}+\sqrt{|\bp_1+\bq-\bp_2-\bp_3|}\,)^{14}\,(\tau_{\rm {E1},1}^B\,\tau_{\rm {E1},2}^B\,\tau_{\rm {E1},3}^B\,\tau_{\rm {E1},4}^B)^{2\pi\delta}}\nonumber\\
		&+\,\frac{q^{1/2}\,|\bp_1+\bq|^{1/2}\,|\bq-\bp_2-\bp_3|^{1/2}\,|\bp_1+\bq-\bp_2-\bp_3|^{1/2}}{(\sqrt{q}+\sqrt{|\bp_1+\bq|}\,)^7\,(\sqrt{q}+\sqrt{|\bq-\bp_2-\bp_3|}\,)^7\,(\sqrt{|\bq-\bp_2-\bp_3|}+\sqrt{|\bp_1+\bq-\bp_2-\bp_3|}\,)^7}\nonumber \\&\times \frac{\mathcal{A}_C(\bq, -\bp_1-\bq, \bq-\bp_2-\bp_3,\bp_1+\bq-\bp_2-\bp_3)}{(\sqrt{|\bp_1+\bq|}+\sqrt{|\bp_1+\bq-\bp_2-\bp_3|}\,)^7\,(\tau_{\rm {E1},1}^C\,\tau_{\rm {E1},2}^C\,\tau_{\rm {E1},3}^C\,\tau_{\rm {E1},4}^C)^{2\pi\delta}}\nonumber\\& + 2\,\frac{p_2^{1/2}\,q^{1/2}\,|\bp_1+\bp_2|^{1/2}\,|\bp_1-\bq|^{1/2}\,\mathcal{A}_E(\bp_2, -\bp_1-\bp_2, \bq,\bp_1-\bq)}{(\sqrt{p_2}+\sqrt{|\bp_1+\bp_2|}\,)^{14}\,(\sqrt{q}+\sqrt{|\bp_1-\bq|}\,)^{14}\,(\tau_{\rm {E1},1}^E\,\tau_{\rm {E1},2}^E\,\tau_{\rm {E1},3}^E\,\tau_{\rm {E1},4}^E)^{2\pi\delta}}\biggr)\,,
	\end{align}
where we have neglected the $k$ terms in the sums and with the time parameters given by
\begin{equation*}
	\renewcommand{\arraystretch}{1.2}
	\setlength{\tabcolsep}{5pt}
	{\fontsize{9.5}{11}
		\begin{array}{l@{\extracolsep{2pt}}l} 
			\tau_{E1,1}^{A} = \tau_{E1,2}^{A} = \tau_{E1,1}^{B} = \tau_{E1,2}^{B} = \tau_{E1,1}^{C} \simeq -(\sqrt{q} + \sqrt{|\mathbf{p}_1 + \mathbf{q}|})^{-2}\,, & \\
			\tau_{E1,3}^{A} = \tau_{E1,4}^{A} \simeq -(\sqrt{|\mathbf{p}_1 + \mathbf{q}|} + \sqrt{|\mathbf{p}_1 + \mathbf{p}_2 + \mathbf{p}_3+\bq|})^{-2}, &\tau_{E1,2}^{C} \simeq -(\sqrt{q} + \sqrt{|\mathbf{q} - \mathbf{p}_2 - \mathbf{p}_3|})^{-2}, \\
			\tau_{E1,3}^{B} = \tau_{E1,4}^{B} \simeq -(\sqrt{|\mathbf{p}_1 + \mathbf{q}|} + \sqrt{|\mathbf{p}_1 + \mathbf{q} -\mathbf{p}_2- \mathbf{p}_3|})^{-2}, &  \tau_{E1,1}^{E} = \tau_{E1,2}^{E} \simeq -(\sqrt{p_2} + \sqrt{|\mathbf{p}_1 + \mathbf{p}_2|})^{-2},\\
			\tau_{E1,3}^{C} \simeq -(\sqrt{|\mathbf{q} - \mathbf{p}_2 - \mathbf{p}_3|} + \sqrt{|\mathbf{p}_1 + \mathbf{q} - \mathbf{p}_2 - \mathbf{p}_3|})^{-2}\,, & \tau_{E1,3}^{E} = \tau_{E1,4}^{E} \simeq -(\sqrt{q} + \sqrt{|\mathbf{p}_1 - \mathbf{q}|})^{-2}\,,\\
			\tau_{E1,4}^{C} \simeq -(\sqrt{|\mathbf{p}_1 + \mathbf{q}|} + \sqrt{|\mathbf{p}_1 + \mathbf{q} - \mathbf{p}_2 - \mathbf{p}_3|})^{-2}. &  \\
	\end{array}}
\end{equation*}
\normalsize

In order to compute \eqref{mixed_I_correlator_explicit_large_momenta}, we again consider specific orderings of the momenta and integrate over one momentum at a time keeping only the dominant terms, as done in Appendix~\ref{appendix_homogeneous}. Many orderings give significantly suppressed results due to the $2^7$ factors in the denominators whenever the largest momentum appears in both roots of a sum. As a result, the number of contributing cases is greatly reduced. For the first expression of \eqref{mixed_I_correlator_explicit_large_momenta}, there are six cases that contribute significantly:\small
\renewcommand{\arraystretch}{2.5}
\begin{center}
		\begin{tabular}{|c|c|}
		\hline
	$p_3>p_2>p_1>q$	& $\dfrac{5.1\times 10^4}{k_{BR}^3\,(4\pi\delta+1/2)\,(4\pi\delta+7/2)\,(8\pi\delta-3)}$ \\[2mm]
		\hline
	$p_2>p_3>p_1>q$	&   $\dfrac{5.1\times 10^4}{k_{BR}^3\,(4\pi\delta+1/2)^2\,(8\pi\delta-3)}$ \\[2mm]
		\hline
	$p_3>p_1>p_2>q$	& $\dfrac{7.8\times 10^3}{k_{BR}^3\,(4\pi\delta+7/2)\,(8\pi\delta-3)}$ \\[2mm]
		\hline
	$p_2>p_1>p_3>q$	&
	 $\dfrac{1.4\times 10^4}{k_{BR}^3\,(4\pi\delta+1/2)\,(8\pi\delta-3)} $\\[2mm]
		\hline
	$p_3>p_1>q>p_2$	& $\dfrac{9.1\times10^3}{k_{BR}^3\,(4\pi\delta+7/2)\,(8\pi\delta-3)}$ \\[2mm]
		\hline
		$p_2>p_1>q>p_3$& $\,\dfrac{ 1.4\times 10^4\,\left(1.2\times10^3\pi^2\delta^2 - 4.7\times10^2\pi\delta - 22.\right)}{k_{BR}^3\,(8 \pi  \delta-3 )^2 (8 \pi  \delta +1)^2}$\\[2mm]
		\hline
	\end{tabular}
\end{center}\normalsize
	valid for $\delta>\frac{3}{8\pi}$.	The second expression of \eqref{mixed_I_correlator_explicit_large_momenta}  has the same dependence on momenta as the first one and so gives the same result. For the third expression of \eqref{mixed_I_correlator_explicit_large_momenta} there are not combinations which do not suffer the suppression. Finally, for the fourth expression, we consider all combinations where $p_1 > q$ and $p_1 > p_2$. Although there are eight such combinations, the symmetry between $p_2$ and $q$ in the integral ensures that swapping $p_2$ and $q$ gives the same result. Therefore, we need to compute only four distinct cases:\small
	\renewcommand{\arraystretch}{2.5}
	\begin{center}
	\begin{tabular}{|c|c|}
		\hline
	$p_1>p_3>q>p_2$ (or $p_1>p_3>p_2>q$)  & $\dfrac{1.3\times 10^2}{k_{BR}^3\,(8\pi\delta-3)}$ \\[2mm]
		\hline
	$p_1>q>p_3>p_2$ (or $p_1>p_2>p_3>q$) 	&   $\dfrac{2.6\times 10^2}{k_{BR}^3\,(8\pi\delta-3)}$ \\[2mm]
		\hline
	$p_1>q>p_2>p_3$ (or $p_1>p_2>q>p_3$) & $\dfrac{1.3\times 10^2 \,(1.4\times 10^2\,\pi\delta - 6.1\times10)}{k_{BR}^3\, (8 \pi  \delta-3 )^2} $\\[2mm]
		\hline
	 $p_3>p_1>q>p_2$ (or $p_3>p_1>p_2>q$) 	&
		$\dfrac{9.\times10^2}{k_{BR}^3\,(8\pi\delta-3)^2} $\\[2mm]
		\hline
	\end{tabular}
\end{center}\normalsize
valid for $\delta>\frac{3}{8\pi}$. Summing all contributions, we obtain
	\begin{align}
	\mathit{I}_{\rm E,1}=\frac{-1.4 \times 10^2 +  1. \times 10^3\,\delta + 4.3 \times 10^2\, \delta^2 }{k_{BR}^3\,\left(1.2 \times 10^{-1} - 1.\, \delta \right)^2 \left(4. \times 10^{-2} + 1.\, \delta \right)} \,,
		\hspace{0.2cm}\text{for} \hspace{0.1cm} \delta > \frac{3}{8\pi} \,.
	\end{align}
	In particular for $\delta= 0.2$ the integral takes the value 	\begin{align}\label{result1}
	\mathit{I}_{\rm E,1}\simeq\dfrac{5.2\times 10^4}{k^3}\,\biggl(\frac{k}{k_{BR}}\biggr)^3\,,\end{align}used in~\eqref{mixed_I_large_momenta}.
	\subsection{Term: $ {\cal C}_{\Omega \Omega,2}^{\rm E}(\bk)$}	
	\label{Term 2}
		The second contribution to the extrinsic correlator of the gravitational wave energy densities in the regime of large momenta, eq.~\eqref{mixed_II_large_momenta}, is found by evaluating the integral
	\begin{align}\label{mixed_II_correlator_explicit_large_momenta}
		&\mathit{I}_{\rm E,2}=\frac{1}{k_{BR}^{8\pi\delta}}\int d\bp_1 \,d\bp_2\,d\bp_3\,d\bq\,\frac{1}{p_3^3} \nonumber\\
		&\times\,\biggl(\frac{q^{1/2}\,|\bp_1+\bp_3+\bq|\,|\bp_1+\bp_2+\bp_3+\bq|^{1/2}}{(\sqrt{q}+\sqrt{|\bp_1+\bp_3+\bq|}\,)^{14}}\nonumber\\&\times\,\frac{\mathcal{A}_A(\bq,-\bp_1-\bp_3-\bq,\bp_1+\bp_3+\bq,\bp_1+\bp_2+\bp_3+\bq)}{(\sqrt{|\bp_1+\bp_3+\bq|}+\sqrt{|\bp_1+\bp_2+\bp_3+\bq|}\,)^{14}\,(\tau_{\rm {E2},1}^A\,\tau_{\rm {E2},2}^A\,\tau_{\rm {E2},3}^A\,\tau_{\rm {E2},4}^A)^{2\pi\delta}}\nonumber \\
		&+\,\frac{q^{1/2}\,|\bp_1+\bp_3+\bq|\,|\bp_1+\bq-\bp_2+\bp_3|^{1/2}}{(\sqrt{q}+\sqrt{|\bp_1+\bp_3+\bq|}\,)^{14}}\nonumber\\&\times \,\frac{\mathcal{A}_B(\bq, -\bp_1-\bp_3-\bq, \bp_1+\bp_3+\bq,\bp_1+\bq-\bp_2+\bp_3)}{(\sqrt{|\bp_1+\bp_3+\bq|}+\sqrt{|\bp_1+\bq-\bp_2+\bp_3|}\,)^{14}\,(\tau_{\rm {E2},1}^B\,\tau_{\rm {E2},2}^B\,\tau_{\rm {E2},3}^B\,\tau_{\rm {E2},4}^B)^{2\pi\delta}}\nonumber\\
		&+\,\frac{q^{1/2}\,|\bp_1+\bp_3+\bq|^{1/2}\,|\bq-\bp_2|^{1/2}\,|\bp_1+\bq-\bp_2+\bp_3|^{1/2}}{(\sqrt{q}+\sqrt{|\bp_1+\bp_3+\bq|}\,)^7\,(\sqrt{q}+\sqrt{|\bq-\bp_2|}\,)^7\,(\sqrt{|\bq-\bp_2|}+\sqrt{|\bp_1+\bq-\bp_2+\bp_3|}\,)^7}\nonumber \\&\times \frac{\mathcal{A}_C(\bq, -\bp_1-\bp_3-\bq, \bq-\bp_2,\bp_1+\bq-\bp_2+\bp_3)}{(\sqrt{|\bp_1+\bp_3+\bq|}+\sqrt{|\bp_1+\bq-\bp_2+\bp_3|}\,)^7\,(\tau_{\rm {E2},1}^C\,\tau_{\rm {E2},2}^C\,\tau_{\rm {E2},3}^C\,\tau_{\rm {E2},4}^C)^{2\pi\delta}}\nonumber\\& + 2\,\frac{p_2^{1/2}\,q^{1/2}\,|\bp_1+\bp_2+\bp_3|^{1/2}\,|\bp_1+\bp_3-\bq|^{1/2}\,\mathcal{A}_E(\bp_2, -\bp_1-\bp_2-\bp_3, \bq,\bp_1+\bp_3-\bq)}{(\sqrt{p_2}+\sqrt{|\bp_1+\bp_2+\bp_3|}\,)^{14}\,(\sqrt{q}+\sqrt{|\bp_1+\bp_3-\bq|}\,)^{14}\,(\tau_{\rm {E2},1}^E\,\tau_{\rm {E2},2}^E\,\tau_{\rm {E2},3}^E\,\tau_{\rm {E2},4}^E))^{2\pi\delta}}\biggr)\,,
	\end{align}	where we have neglected the $k$ terms in the sums and with the time parameters given by
\small
  {\begin{align*}
			&\tau_{\rm {E2},1}^A =\tau_{\rm {E2},2}^A=\tau_{\rm {E2},1}^B = \tau_{\rm {E2},2}^B= \tau_{\rm {E2},1}^C\simeq - (\sqrt{q}+\sqrt{|\bp_1+\bp_3+\bq|}\,)^{-2}\,,\nonumber \\&\tau_{\rm {E2},3}^A=\tau_{\rm {E2},4}^A \simeq - (\sqrt{|\bp_1+\bp_3+\bq|}+\sqrt{|\bp_1+\bp_2+\bp_3+\bq|}\,)^{-2}\,, \nonumber \\&
		\tau_{\rm {E2},3}^B=\tau_{\rm {E2},4}^B \simeq - (\sqrt{|\bp_1+\bp_3+\bq|}+\sqrt{|\bp_1+\bq -\bp_2+\bp_3|}\,)^{-2}\,, \nonumber \\
		&
		\tau_{\rm {E2},2}^C\simeq - (\sqrt{q}+\sqrt{|\bq-\bp_2|}\,)^{-2}\,, \nonumber \\
		&
		\tau_{\rm {E2},3}^C\simeq - (\sqrt{|\bq-\bp_2|}+\sqrt{|\bp_1+\bq-\bp_2+\bp_3|}\,)^{-2}\,, \nonumber \\&
		\tau_{\rm {E2},4}^C\simeq - (\sqrt{|\bp_1+\bp_3+\bq|}+\sqrt{|\bp_1+\bq-\bp_2+\bp_3|}\,)^{-2}\,, \nonumber \\&
		\tau_{\rm {E2},1}^E=\tau_{\rm {E2},2}^E \simeq - (\sqrt{p_2}+\sqrt{|\bp_1+\bp_2+\bp_3|}\,)^{-2}\,, \nonumber \\&
		\tau_{\rm {E2},3}^E=\tau_{\rm {E2},4}^E\simeq - (\sqrt{q}+\sqrt{|\bp_1+\bp_3-\bq|}\,)^{-2}\,.
	\end{align*}}
\normalsize

	We perform again one integration at the time considering all the possible orderings of the four momenta, i.e. $4!= 24$ permutations, but neglecting all the terms suppressed by  $2^7$ factors at the denominators, as done in Appendix~\ref{appendix_homogeneous}. For the first expression of \eqref{mixed_II_correlator_explicit_large_momenta} the only cases that survive are the four cases with $q< p_1 \text{ or } p_3$ and $p_2> p_1, p_3 \text{ and } q$: \small
	\renewcommand{\arraystretch}{2.5}
	\begin{center}
	\begin{tabular}{|c|c|}
		\hline
		$p_2>p_1>p_3>q$	& $\dfrac{1.4\times 10^4}{k_{BR}^3\,(4\pi\delta+1/2)\,(8\pi\delta-3)}$ \\[2mm]
		\hline
$p_2>p_3>p_1>q$	&   $\dfrac{7.8\times10^3}{k_{BR}^3\,(4\pi\delta+1/2)\,(8\pi\delta-3)}  $ \\[2mm]
		\hline
	$p_2>p_1>q>p_3$	& $\dfrac{ 1.4\times10^4\,\left(1.2\times10^3\pi^2\delta^2 - 4.7\times10^2\pi\delta - 22.\right)}{k_{BR}^3\, (8 \pi  \delta-3 )^2 (8 \pi  \delta +1)^2} $ \\[2mm]
		\hline
	$p_2>p_3>q>p_1$	& $\dfrac{9.1\times10^3}{k_{BR}^3\,(4\pi\delta+1/2)\,(8\pi\delta-3)} $\\[2mm]
		\hline
	\end{tabular}
\end{center}\normalsize
			valid for $\delta>\frac{3}{8\pi}$.	The second expression of \eqref{mixed_II_correlator_explicit_large_momenta} has the same dependence on momenta as the first one and so gives the same result. The third expression contains only suppressed combinations. Finally, for the fourth expression, we consider all combinations where $p_1 \text{ or } p_3> q$ and $p_1 \text{ or } p_3> p_2$. Although there are twelve such combinations, the symmetry between $p_2$ and $q$ allow us to compute only six of them:\small
	\renewcommand{\arraystretch}{2.5}
	\begin{center}
	\begin{tabular}{|c|c|}
		\hline
	$p_1>p_3>q>p_2$ (or $p_1>p_3>p_2>q$) 	& $\dfrac{1.3\times10^2}{k_{BR}^3\,(8\pi\delta-3)}$ \\[2mm]
		\hline
	$p_1>q>p_3>p_2$ (or $p_1>p_2>p_3>q$) 	&   $  \dfrac{2.6\times10^2}{k_{BR}^3\,(8\pi\delta-3)}$ \\[2mm]
		\hline
	$p_1>q>p_2>p_3$ (or $p_1>p_2>q>p_3$) 	& $\dfrac{1.3\times10^2 \,(1.4\times 10^2\,\pi\delta - 6.1\times10))}{k_{BR}^3\, (8 \pi  \delta -3)^2}$ \\[2mm]
		\hline
	 $p_3>p_1>q>p_2$ (or $p_3>p_1>p_2>q$) 	&	$\dfrac{9.\times10^1}{k_{BR}^3\,(8\pi\delta-3)}$\\[2mm]
		\hline
	 $p_3>q>p_1>p_2$ (or $p_3>p_2>p_1>q$) 	& $\dfrac{9.7\times10^1}{k_{BR}^3\,(8\pi\delta-3)}$ \\[2mm]
		\hline
	$p_3>q>p_2>p_1$ (or $p_3>p_2>q>p_1$)	& $\dfrac{1.1\times10^2}{k_{BR}^3\,(8\pi\delta-3)}$\\[2mm]
		\hline
	\end{tabular}
\end{center}\normalsize
	valid for $\delta>\frac{3}{8\pi}$. Summing all contributions, we obtain
	\begin{align}
		\mathit{I}_{\rm E,2}= \frac{-2.6 - 1.3 \times 10^2\, \delta + 1. \times 10^3\, \delta^2 + 4.8 \times 10^2\, \delta^3}
		{k_{BR}^3\,\left(4.7 \times 10^{-3} + 8. \times 10^{-2}\, \delta - 1. \,\delta^2\right)^2}\,,
		\hspace{0.2cm}\text{for} \hspace{0.1cm} \delta > \frac{3}{8\pi} \,.
	\end{align}
	In particular for $\delta= 0.2$ the integral takes the value  \begin{align}\label{result2}
	\mathit{I}_{\rm E,2}\simeq\dfrac{4.7\times 10^4}{k^3}\,\biggl(\frac{k}{k_{BR}}\biggr)^3\,,\end{align}used in~\eqref{mixed_II_large_momenta}.
	\subsection{Term: $ {\cal C}_{\Omega \Omega,3}^{\rm E}(\bk)'$}
	\label{Term 3: Non scale-invariant part}
		The non scale-invariant part of the third contribution to the extrinsic correlator of the gravitational wave energy densities in the regime of large momenta, eq.~\eqref{mixed_III_large_momenta}, is found by evaluating the integral
	\begin{align}\label{mixed_III_correlator_explicit_large_momenta}
		&\mathit{I}_{\rm E,3}=\frac{1}{k_{BR}^{8\pi\delta}}\int d\bp_1 \,d\bp_2\,d\bp_3\,d\bq\,\frac{1}{p_3^3} \nonumber\\
		&\times\,\biggl(\frac{q^{1/2}\,|\bp_1+\bq|^{1/2}\,|\bp_1-\bp_3+\bq|^{1/2}\,|\bp_1+\bp_2+\bq|^{1/2}}{(\sqrt{q}+\sqrt{|\bp_1+\bq|}\,)^{7}\,(\sqrt{q}+\sqrt{|\bp_1-\bp_3+\bq|}\,)^{7}\,(\sqrt{|\bp_1+\bq|}+\sqrt{|\bp_1+\bp_2+\bq|}\,)^{7}}\nonumber\\&\times\,\frac{\mathcal{A}_A(\bq,-\bp_1-\bq, \bp_1-\bp_3+\bq,\bp_1+\bp_2+\bq)}{(\sqrt{|\bp_1-\bp_3+\bq|}+\sqrt{|\bp_1+\bp_2+\bq|}\,)^{7}\,(\tau_{\rm {E3},1}^A\,\tau_{\rm {E3},2}^A\,\tau_{\rm {E3},3}^A\,\tau_{\rm {E3},4}^A)^{2\pi\delta}}\nonumber \\
		&+\,\frac{q^{1/2}\,|\bp_1+\bq|^{1/2}\,|\bp_1-\bp_3+\bq|^{1/2}\,|\bp_1-\bp_3-\bp_2+\bq|^{1/2}}{(\sqrt{q}+\sqrt{|\bp_1+\bq|}\,)^{7}\,(\sqrt{q}+\sqrt{|\bp_1-\bp_3+\bq|}\,)^{7}\,(\sqrt{|\bp_1+\bq|}+\sqrt{|\bp_1-\bp_3-\bp_2+\bq|}\,)^{7}}\nonumber\\&\times\,\frac{\mathcal{A}_B(\bq,-\bp_1-\bq, \bp_1-\bp_3+\bq,\bp_1-\bp_3-\bp_2+\bq)}{(\sqrt{|\bp_1-\bp_3+\bq|}+\sqrt{|\bp_1-\bp_3-\bp_2+\bq|}\,)^{7}\,(\tau_{\rm {E3},1}^B\,\tau_{\rm {E3},2}^B\,\tau_{\rm {E3},3}^B\,\tau_{\rm {E3},4}^B)^{2\pi\delta}}\nonumber \\
		&+\,\frac{q^{1/2}\,|\bp_1+\bq|^{1/2}\,|\bq-\bp_2|^{1/2}\,|\bp_1-\bp_3-\bp_2+\bq|^{1/2}}{(\sqrt{q}+\sqrt{|\bp_1+\bq|}\,)^7\,(\sqrt{q}+\sqrt{|\bq-\bp_2|}\,)^7\,(\sqrt{|\bq-\bp_2|}+\sqrt{|\bp_1-\bp_3-\bp_2+\bq|}\,)^7}\nonumber \\&\times \frac{\mathcal{A}_C(\bq, -\bp_1-\bq, \bq-\bp_2,\bp_1-\bp_3-\bp_2+\bq)}{(\sqrt{|\bp_1+\bq|}+\sqrt{|\bp_1-\bp_3-\bp_2+\bq|}\,)^7\,(\tau_{\rm {E3},1}^C\,\tau_{\rm {E3},2}^C\,\tau_{\rm {E3},3}^C\,\tau_{\rm {E3},4}^C)^{2\pi\delta}}\nonumber\\& + 2\,\frac{p_2^{1/2}\,q^{1/2}\,|\bp_1+\bp_2|^{1/2}\,|\bp_1-\bp_3-\bq|^{1/2}\,\mathcal{A}_E(\bp_2, -\bp_1-\bp_2, \bq,\bp_1-\bp_3-\bq)}{(\sqrt{p_2}+\sqrt{|\bp_1+\bp_2|}\,)^{14}\,(\sqrt{q}+\sqrt{|\bp_1-\bp_3-\bq|}\,)^{14}\,(\tau_{\rm {E3},1}^E\,\tau_{\rm {E3},2}^E\,\tau_{\rm {E3},3}^E\,\tau_{\rm {E3},4}^E)^{2\pi\delta}}\biggr)\,,
	\end{align}
where we have neglected the $k$ terms in the sums and with the time parameters given by
	\begin{equation*}
		\renewcommand{\arraystretch}{1.2}
		\setlength{\tabcolsep}{5pt}
		{\fontsize{9.5}{11}
			\begin{array}{l@{\extracolsep{5pt}}l} 
				\tau_{\rm {E3},1}^A = \tau_{\rm {E3},1}^B = \tau_{\rm {E3},1}^C \simeq - (\sqrt{q}+\sqrt{|\bp_1+\bq|}\,)^{-2}\,, & \tau_{\rm {E3},3}^B \simeq - (\sqrt{|\bp_1+\bq|}+\sqrt{|\bp_1-\bp_3-\bp_2+\bq|}\,)^{-2}\,, \\
				\tau_{\rm {E3},2}^A = \tau_{\rm {E3},2}^B \simeq - (\sqrt{q}+\sqrt{|\bp_1-\bp_3+\bq|}\,)^{-2}\,, & \tau_{\rm {E3},2}^C \simeq - (\sqrt{q}+\sqrt{|\bq-\bp_2|}\,)^{-2}\,, \\
				\tau_{\rm {E3},3}^A \simeq - (\sqrt{|\bp_1+\bq|}+\sqrt{|\bp_1+\bp_2+\bq|}\,)^{-2}\,, & \tau_{\rm {E3},3}^C \simeq - (\sqrt{|\bq-\bp_2|}+\sqrt{|\bp_1-\bp_3-\bp_2+\bq|}\,)^{-2}
				\,, \\
				\tau_{\rm {E3},4}^A \simeq - (\sqrt{|\bp_1-\bp_3+\bq|}+\sqrt{|\bp_1+\bp_2+\bq|}\,)^{-2}\,, & 
				\tau_{\rm {E3},1}^E = \tau_{\rm {E3},2}^E \simeq - (\sqrt{p_2}+\sqrt{|\bp_1+\bp_2|}\,)^{-2}
				\,, \\
				\tau_{\rm {E3},4}^B \simeq - (\sqrt{|\bp_1-\bp_3+\bq|}+\sqrt{|\bp_1-\bp_3-\bp_2+\bq|}\,)^{-2}\,, & 
					\tau_{\rm {E3},3}^E = \tau_{\rm {E3},4}^E \simeq - (\sqrt{q}+\sqrt{|\bp_1-\bp_3-\bq|}\,)^{-2}\,, \\
				\tau_{\rm {E3},4}^C \simeq - (\sqrt{|\bp_1+\bq|}+\sqrt{|\bp_1-\bp_3-\bp_2+\bq|}\,)^{-2}\,. & \\
		\end{array}}
	\end{equation*}
\normalsize

	We perform again one integration at the time neglecting all the terms suppressed by  $2^7$ factors, as done in Appendix~\ref{appendix_homogeneous}. For the first expression of \eqref{mixed_III_correlator_explicit_large_momenta} the only cases that survive are the four cases with $p_2>p_1>q$: \small
\renewcommand{\arraystretch}{2.5}
\begin{center}
\begin{tabular}{|c|c|}
	\hline
$p_2>p_1>q>p_3$	& $\dfrac{1.4\times10^4\left(1.2\times10^3\pi^2\delta^2 - 4.7\times10^2\pi\delta - 22.\right)}{k_{BR}^3\,(8 \pi  \delta -3)^2 (8 \pi  \delta +1)^2}$ \\[2mm]
	\hline
	 $p_2>p_1>p_3>q$&   $\dfrac{1.4\times10^4}{k_{BR}^3\,(4\pi\delta+1/2)\,(8\pi\delta-3)}$ \\[2mm]
	\hline
 $p_2>p_3>p_1>q$	& $\dfrac{1.7\times10^4}{k_{BR}^3\,(2\pi\delta+7/2)\,(4\pi\delta+1/2)\,(8\pi\delta-3)} $ \\[2mm]
	\hline
$p_3>p_2>p_1>q$	&	$ \dfrac{1.7\times10^4}{k_{BR}^3\,(2\pi\delta+7/2)\,(4\pi\delta+7/2)\,(8\pi\delta-3)}$\\[2mm]
	\hline
\end{tabular}
\end{center}\normalsize
	valid for $\delta>\frac{3}{8\pi}$. For the second expression, the cases that contribute are those where $p_2$ is the largest among all momenta, with $p_1 > q$. This condition leaves us with three possible cases:\small
	\renewcommand{\arraystretch}{2.5}
	\begin{center}
	\begin{tabular}{|c|c|}
		\hline
	 $p_2>p_3>p_1>q$  	& $\dfrac{1.7\times10^4}{k_{BR}^3\,(2\pi\delta+7/2)\,(4\pi\delta+1/2)\,(8\pi\delta-3)}$ \\[2mm]
		\hline
	$p_2>p_1>p_3>q$ 	&   $\dfrac{1.4\times10^4}{k_{BR}^3\,(4\pi\delta+1/2)\,(8\pi\delta-3)} $ \\[2mm]
		\hline
	$p_2>p_1>q>p_3$ 	& $\dfrac{1.4\times10^4\left(1.2\times10^3\pi^2\delta^2 - 4.7\times10^2\pi\delta - 22.\right)}{k_{BR}^3\,(8 \pi  \delta-3 )^2 (8 \pi  \delta +1)^2}$ \\[2mm]
		\hline
	\end{tabular}
\end{center}\normalsize
	valid for $\delta>\frac{3}{8\pi}$.	In the third expression, only two cases survive, i.e. $p_3 > p_2 > p_1 > q$ and $p_3 > p_1 > p_2 > q$, which give rise to the same result:
	\begin{align}
 \frac{6.3\times10^3}{k_{BR}^3\,(2\pi\delta+7/2)\,(4\pi\delta+7/2)\,(8\pi\delta-3)}\,,\hspace{0.4cm}\text{for} \hspace{0.2cm}\delta>\frac{3}{8\pi}\,.
	\end{align}
	In the fourth expression, we have nine cases, satisfying $p_1>p_2$ and $q< p_1 \text{ or } p_3$, which reduce to  five if we use the symmetry under the exchange $p_2 \leftrightarrow q$:\small
	\renewcommand{\arraystretch}{2.5}
		\begin{center}
	\begin{tabular}{|c|c|}
		\hline
	$p_1>p_2>q>p_3$ (or $p_1>q>p_2>p_3$) 	& $\dfrac{1.3\times10^2 \,(1.4\times 10^2\,\pi\delta - 6.1\times10)}{k_{BR}^3\, (8 \pi  \delta -3)^2}$ \\[2mm]
		\hline
	 $p_1>p_2>p_3>q$ (or $p_1>q>p_3>p_2$) 	&   $\dfrac{2.6\times10^2}{k_{BR}^3\,(8\pi\delta-3)} $ \\[2mm]
		\hline
	$p_1>p_3>p_2>q$ (or $p_1>p_3>q>p_2$)	& $ \dfrac{1.3\times10^2}{k_{BR}^3\,(8\pi\delta-3)}$ \\[2mm]
		\hline
	$p_3>p_1>p_2>q$ (or $p_3>p_1>q>p_2$ ) 	&	$\dfrac{3.\times10^2}{k_{BR}^3\,(4\pi\delta+7/2)\,(8\pi\delta-3)} $\\[2mm]
		\hline
	$p_3>q>p_1>p_2$  	& $\dfrac{2.1\times10^3}{k_{BR}^3\,(4\pi\delta)\,(4\pi\delta+7/2)\,(8\pi\delta-3)}  $ \\[2mm]
		\hline
	\end{tabular}
\end{center}\normalsize
	valid for $\delta>\frac{3}{8\pi}$. Summing all contributions, we obtain
	\begin{align}
	\mathit{I}_{\rm E,3}&=\frac{1}{k_{BR}^3\,\delta \left(2.8 \times 10^{-1} + 1.\, \delta \right) 
		\left(5.6 \times 10^{-1} + 1.\, \delta \right) 
		\left(4.7 \times 10^{-3} + 8. \times 10^{-2}\,\delta - 1. \,\delta^2\right)^2
	}\nonumber\\& \times
		(-5.6 \times 10^{-5} - 3.5 \times 10^{-1}\,\delta - 2. \times 10^1\,\delta^2 
		+ 5.1 \times 10^1\,\delta^3 + 7.4 \times 10^2\,\delta^4 
		 \nonumber\\& + 1.1 \times 10^3\,\delta^5+ 2.2 \times 10^2\,\delta^6)\,,
\end{align}
	valid for $\delta>\frac{3}{8\pi}$. 	In particular for $\delta= 0.2$ the integral takes the value 	\begin{align}\label{result3}
	\mathit{I}_{\rm E,3}=\dfrac{4.\times 10^4}{k^3}\biggl(\frac{k}{k_{BR}}\biggr)^3\,,\end{align}used in~\eqref{mixed_III_large_momenta}.


\begin{thebibliography}{99}
		
		\bibitem{LIGOScientific:2016aoc}
		B.~P.~Abbott \textit{et al.} [LIGO Scientific and Virgo],
		Phys. Rev. Lett. \textbf{116}, no.6, 061102 (2016)
		doi:10.1103/PhysRevLett.116.061102
		[arXiv:1602.03837 [gr-qc]].
		
		\bibitem{NANOGrav:2023gor}
		G.~Agazie \textit{et al.} [NANOGrav],
		Astrophys. J. Lett. \textbf{951}, no.1, L8 (2023)
		doi:10.3847/2041-8213/acdac6
		[arXiv:2306.16213 [astro-ph.HE]].
		
		\bibitem{EPTA:2023fyk}
		J.~Antoniadis \textit{et al.} [EPTA and InPTA:],
		``The second data release from the European Pulsar Timing Array - III. Search for gravitational wave signals,''
		Astron. Astrophys. \textbf{678}, A50 (2023)
		doi:10.1051/0004-6361/202346844
		[arXiv:2306.16214 [astro-ph.HE]].
		
		\bibitem{Reardon:2023gzh}
		D.~J.~Reardon, A.~Zic, R.~M.~Shannon, G.~B.~Hobbs, M.~Bailes, V.~Di Marco, A.~Kapur, A.~F.~Rogers, E.~Thrane and J.~Askew, \textit{et al.}
		``Search for an Isotropic Gravitational-wave Background with the Parkes Pulsar Timing Array,''
		Astrophys. J. Lett. \textbf{951}, no.1, L6 (2023)
		doi:10.3847/2041-8213/acdd02
		[arXiv:2306.16215 [astro-ph.HE]].
		
		\bibitem{Allen:1996vm}
		B.~Allen,
		[arXiv:gr-qc/9604033 [gr-qc]].
		
		\bibitem{Caprini:2018mtu}
		C.~Caprini and D.~G.~Figueroa,
		Class. Quant. Grav. \textbf{35}, no.16, 163001 (2018)
		doi:10.1088/1361-6382/aac608
		[arXiv:1801.04268 [astro-ph.CO]].

		\bibitem{Regimbau:2011rp}
		T.~Regimbau,
		Res. Astron. Astrophys. \textbf{11}, 369-390 (2011)
		doi:10.1088/1674-4527/11/4/001
		[arXiv:1101.2762 [astro-ph.CO]].
		
		\bibitem{Pajer:2013fsa}
		E.~Pajer and M.~Peloso,
		``A review of Axion Inflation in the era of Planck,''
		Class. Quant. Grav. \textbf{30}, 214002 (2013)
		doi:10.1088/0264-9381/30/21/214002
		[arXiv:1305.3557 [hep-th]].
		
		\bibitem{Freese:1990rb}
		K.~Freese, J.~A.~Frieman and A.~V.~Olinto,
		``Natural inflation with pseudo - Nambu-Goldstone bosons,''
		Phys. Rev. Lett. \textbf{65} (1990), 3233-3236
		doi:10.1103/PhysRevLett.65.3233
		
		\bibitem{Sorbo:2011rz}
		L.~Sorbo,
		``Parity violation in the Cosmic Microwave Background from a pseudoscalar inflaton,''
		JCAP \textbf{06}, 003 (2011)
		doi:10.1088/1475-7516/2011/06/003
		[arXiv:1101.1525 [astro-ph.CO]].
		
		\bibitem{Barnaby:2010vf}
		N.~Barnaby and M.~Peloso,
		``Large Nongaussianity in Axion Inflation,''
		Phys. Rev. Lett. \textbf{106}, 181301 (2011)
		doi:10.1103/PhysRevLett.106.181301
		[arXiv:1011.1500 [hep-ph]].
		
		\bibitem{Namba:2015gja}
		R.~Namba, M.~Peloso, M.~Shiraishi, L.~Sorbo and C.~Unal,
		``Scale-dependent gravitational waves from a rolling axion,''
		JCAP \textbf{01}, 041 (2016)
		doi:10.1088/1475-7516/2016/01/041
		[arXiv:1509.07521 [astro-ph.CO]].
		
		\bibitem{Linde:2012bt}
		A.~Linde, S.~Mooij and E.~Pajer,
		``Gauge field production in supergravity inflation: Local non-Gaussianity and primordial black holes,''
		Phys. Rev. D \textbf{87} (2013) no.10, 103506
		doi:10.1103/PhysRevD.87.103506
		[arXiv:1212.1693 [hep-th]].
		
		\bibitem{Anber:2015yca}
		M.~M.~Anber and E.~Sabancilar,
		``Hypermagnetic Fields and Baryon Asymmetry from Pseudoscalar Inflation,''
		Phys. Rev. D \textbf{92}, no.10, 101501 (2015)
		doi:10.1103/PhysRevD.92.101501
		[arXiv:1507.00744 [hep-th]].
		
		\bibitem{Garretson:1992vt}
		W.~D.~Garretson, G.~B.~Field and S.~M.~Carroll,
		``Primordial magnetic fields from pseudoGoldstone bosons,''
		Phys. Rev. D \textbf{46}, 5346-5351 (1992)
		doi:10.1103/PhysRevD.46.5346
		[arXiv:hep-ph/9209238 [hep-ph]].
		
		\bibitem{Anber:2006xt}
		M.~M.~Anber and L.~Sorbo,
		JCAP \textbf{10}, 018 (2006)
		doi:10.1088/1475-7516/2006/10/018
		[arXiv:astro-ph/0606534 [astro-ph]].
		
   	    \bibitem{Cook:2011hg}
		J.~L.~Cook and L.~Sorbo,
		``Particle production during inflation and gravitational waves detectable by ground-based interferometers,''
		Phys. Rev. D \textbf{85}, 023534 (2012)
		[erratum: Phys. Rev. D \textbf{86}, 069901 (2012)]
		doi:10.1103/PhysRevD.85.023534
		[arXiv:1109.0022 [astro-ph.CO]].
		
		\bibitem{KAGRA:2021mth}
		R.~Abbott \textit{et al.} [KAGRA, Virgo and LIGO Scientific],
		``Search for anisotropic gravitational-wave backgrounds using data from Advanced LIGO and Advanced Virgo\textquoteright{}s first three observing runs,''
		Phys. Rev. D \textbf{104}, no.2, 022005 (2021)
		doi:10.1103/PhysRevD.104.022005
		[arXiv:2103.08520 [gr-qc]].
		
		\bibitem{LISACosmologyWorkingGroup:2022kbp}
		N.~Bartolo \textit{et al.} [LISA Cosmology Working Group],
		``Probing anisotropies of the Stochastic Gravitational Wave Background with LISA,''
		JCAP \textbf{11}, 009 (2022)
		doi:10.1088/1475-7516/2022/11/009
		[arXiv:2201.08782 [astro-ph.CO]].
		
		\bibitem{Geller:2018mwu}
		M.~Geller, A.~Hook, R.~Sundrum and Y.~Tsai,
		``Primordial Anisotropies in the Gravitational Wave Background from Cosmological Phase Transitions,''
		Phys. Rev. Lett. \textbf{121} (2018) no.20, 201303
		doi:10.1103/PhysRevLett.121.201303
		[arXiv:1803.10780 [hep-ph]].
		
		\bibitem{Malhotra:2020ket}
		A.~Malhotra, E.~Dimastrogiovanni, M.~Fasiello and M.~Shiraishi,
		JCAP \textbf{03}, 088 (2021)
		doi:10.1088/1475-7516/2021/03/088
		[arXiv:2012.03498 [astro-ph.CO]].
		
		\bibitem{Adshead:2020bji}
		P.~Adshead, N.~Afshordi, E.~Dimastrogiovanni, M.~Fasiello, E.~A.~Lim and G.~Tasinato,
		``Multimessenger cosmology: Correlating cosmic microwave background and stochastic gravitational wave background measurements,''
		Phys. Rev. D \textbf{103}, no.2, 023532 (2021)
		doi:10.1103/PhysRevD.103.023532
		[arXiv:2004.06619 [astro-ph.CO]].
		
		\bibitem{Ricciardone:2021kel}
		A.~Ricciardone, L.~V.~Dall'Armi, N.~Bartolo, D.~Bertacca, M.~Liguori and S.~Matarrese,
		``Cross-Correlating Astrophysical and Cosmological Gravitational Wave Backgrounds with the Cosmic Microwave Background,''
		Phys. Rev. Lett. \textbf{127} (2021) no.27, 271301
		doi:10.1103/PhysRevLett.127.271301
		[arXiv:2106.02591 [astro-ph.CO]].
		
		\bibitem{Braglia:2021fxn}
		M.~Braglia and S.~Kuroyanagi,
		``Probing prerecombination physics by the cross-correlation of stochastic gravitational waves and CMB anisotropies,''
		Phys. Rev. D \textbf{104}, no.12, 123547 (2021)
		doi:10.1103/PhysRevD.104.123547
		[arXiv:2106.03786 [astro-ph.CO]].
		
		\bibitem{Schulze:2023ich}
		F.~Schulze, L.~Valbusa Dall'Armi, J.~Lesgourgues, A.~Ricciardone, N.~Bartolo, D.~Bertacca, C.~Fidler and S.~Matarrese,
		``GW\_CLASS: Cosmological Gravitational Wave Background in the Cosmic Linear Anisotropy Solving System,''
		[arXiv:2305.01602 [gr-qc]].
		
          \bibitem{Corba_2024}
          S.~P. Corb\`a and L. Sorbo, "Correlated scalar perturbations and gravitational waves from axion inflation," \emph{Journal of Cosmology and Astroparticle Physics}, vol. 2024, no. 10, p. 024, Oct. 2024. doi:10.1088/1475-7516/2024/10/024 [	arXiv:2403.03338 [astro-ph.CO]].
		
		\bibitem{Mentasti:2023icu}
		G.~Mentasti, C.~R.~Contaldi and M.~Peloso,
		``Intrinsic Limits on the Detection of the Anisotropies of the Stochastic Gravitational Wave Background,''
		Phys. Rev. Lett. \textbf{131}, no.22, 221403 (2023)
		doi:10.1103/PhysRevLett.131.221403
		[arXiv:2301.08074 [gr-qc]].
		
		\bibitem{Cui:2023dlo}
		Y.~Cui, S.~Kumar, R.~Sundrum and Y.~Tsai,
		``Unraveling cosmological anisotropies within stochastic gravitational wave backgrounds,''
		JCAP \textbf{10}, 064 (2023)
		doi:10.1088/1475-7516/2023/10/064
		[arXiv:2307.10360 [astro-ph.CO]].
		
		\bibitem{Bethke:2013aba}
		L.~Bethke, D.~G.~Figueroa and A.~Rajantie,
		``Anisotropies in the Gravitational Wave Background from Preheating,''
		Phys. Rev. Lett. \textbf{111}, no.1, 011301 (2013)
		doi:10.1103/PhysRevLett.111.011301
		[arXiv:1304.2657 [astro-ph.CO]].
		
		\bibitem{Yu:2025jgx}
		Y.~H.~Yu and S.~Wang,
		[arXiv:2504.07838 [astro-ph.CO]].
			
\bibitem{Bartolo:2019oiq}
N.~Bartolo, D.~Bertacca, S.~Matarrese, M.~Peloso, A.~Ricciardone, A.~Riotto and G.~Tasinato,
Phys. Rev. D \textbf{100}, no.12, 121501 (2019)
doi:10.1103/PhysRevD.100.121501
[arXiv:1908.00527 [astro-ph.CO]].
		
		\bibitem{Bartolo:2019yeu}
		N.~Bartolo, D.~Bertacca, S.~Matarrese, M.~Peloso, A.~Ricciardone, A.~Riotto and G.~Tasinato,
		``Characterizing the cosmological gravitational wave background: Anisotropies and non-Gaussianity,''
		Phys. Rev. D \textbf{102}, no.2, 023527 (2020)
		doi:10.1103/PhysRevD.102.023527
		[arXiv:1912.09433 [astro-ph.CO]].
		
		\bibitem{Cheng:2015oqa} 
		S.~L.~Cheng, W.~Lee and K.~W.~Ng, ``Numerical study of pseudoscalar inflation with an axion-gauge field coupling,'' Phys.\ Rev.\ D {\bf 93}, no.
		6, 063510 (2016) [arXiv:1508.00251 [astro-ph.CO]].
		
		\bibitem{Notari:2016npn} 
		A.~Notari and K.~Tywoniuk,``Dissipative Axial Inflation,'' JCAP {\bf 1612}, 038 (2016) [arXiv:1608.06223 [hep-th]].
		
		\bibitem{Sobol:2019xls}
		O.~O.~Sobol, E.~V.~Gorbar and S.~I.~Vilchinskii,
		``Backreaction of electromagnetic fields and the Schwinger effect in pseudoscalar inflation magnetogenesis,''
		Phys. Rev. D \textbf{100}, no.6, 063523 (2019)
		[arXiv:1907.10443 [astro-ph.CO]].
		
		\bibitem{DallAgata:2019yrr} 
		G.~Dall'Agata, S.~Gonz\'alez-Mart\'\i{}n, A.~Papageorgiou and M.~Peloso, ``Warm dark energy,'' JCAP \textbf{08}, 032 (2020)  [arXiv:1912.09950 [hep-th]]. 
		
		\bibitem{Domcke:2020zez}
		V.~Domcke, V.~Guidetti, Y.~Welling and A.~Westphal, ``Resonant backreaction in axion inflation,''
		JCAP \textbf{09}, 009 (2020) [arXiv:2002.02952 [astro-ph.CO]].
		
		\bibitem{Caravano:2022epk}
		A.~Caravano, E.~Komatsu, K.~D.~Lozanov and J.~Weller,
		``Lattice simulations of axion-U(1) inflation,''
		Phys. Rev. D \textbf{108}, no.4, 043504 (2023)
		doi:10.1103/PhysRevD.108.043504
		[arXiv:2204.12874 [astro-ph.CO]].
		
		\bibitem{Peloso:2022ovc}
		M.~Peloso and L.~Sorbo,
		``Instability in axion inflation with strong backreaction from gauge modes,''
		JCAP \textbf{01}, 038 (2023)
		doi:10.1088/1475-7516/2023/01/038
		[arXiv:2209.08131 [astro-ph.CO]].
		
		\bibitem{Figueroa:2023oxc}
		D.~G.~Figueroa, J.~Lizarraga, A.~Urio and J.~Urrestilla,
		``Strong Backreaction Regime in Axion Inflation,''
		Phys. Rev. Lett. \textbf{131}, no.15, 151003 (2023)
		doi:10.1103/PhysRevLett.131.151003
		[arXiv:2303.17436 [astro-ph.CO]].
		
		\bibitem{Garcia-Bellido:2023ser}
		J.~Garcia-Bellido, A.~Papageorgiou, M.~Peloso and L.~Sorbo,
		``A flashing beacon in axion inflation: recurring bursts of gravitational waves in the strong backreaction regime,''
		[arXiv:2303.13425 [astro-ph.CO]].
		
		\bibitem{vonEckardstein:2023gwk}
		R.~von Eckardstein, M.~Peloso, K.~Schmitz, O.~Sobol and L.~Sorbo,
		``Axion inflation in the strong-backreaction regime: decay of the Anber-Sorbo solution,''
		JHEP \textbf{11}, 183 (2023)
		doi:10.1007/JHEP11(2023)183
		[arXiv:2309.04254 [hep-ph]].
		
		\bibitem{Caravano:2024xsb}
		A.~Caravano and M.~Peloso,
		[arXiv:2407.13405 [astro-ph.CO]].
		
		\bibitem{Maldacena:2002vr}
		J.~M.~Maldacena,
		``Non-Gaussian features of primordial fluctuations in single field inflationary models,''
		JHEP \textbf{05}, 013 (2003)
		doi:10.1088/1126-6708/2003/05/013
		[arXiv:astro-ph/0210603 [astro-ph]].
		
		\bibitem{Barnaby:2011vw}
		N.~Barnaby, R.~Namba and M.~Peloso,
		``Phenomenology of a Pseudo-Scalar Inflaton: Naturally Large Nongaussianity,''
		JCAP \textbf{04}, 009 (2011)
		doi:10.1088/1475-7516/2011/04/009
		[arXiv:1102.4333 [astro-ph.CO]].
				\end{thebibliography}
\end{document}